\documentclass[aps,floats,superscriptaddress]{revtex4}
 \usepackage{amssymb}
    \usepackage{latexsym}
\usepackage[dvips]{graphicx}

\newcommand{\bc}{\begin{center}}
\newcommand{\ec}{\end{center}}
\def\ba#1{\begin{array}{#1}\displaystyle}
\newcommand{\ea}{\end{array}}

\newcommand{\beq}{\begin{equation}}
\newcommand{\eeq}{\end{equation}}
\newcommand{\beqa}{\begin{eqnarray}}
\newcommand{\eeqa}{\end{eqnarray}}
\newcommand{\no}{\nonumber}
\newcommand{\n}{\nonumber\\}
\newcommand{\bi}{\begin{itemize}}
\newcommand{\ei}{\end{itemize}}
\def\mato#1{\left(\ba{#1}} 
\def\matf{\ea\right)}

\def\sect#1{\section{#1}}

\def\lt#1{\left#1}
\def\rt#1{\right#1}

\def\frc#1#2{\frac{#1}{#2}}

\newcommand{\prin}{\underline{\mathrm{P}}}

\newcommand{\p}{\partial}
\newcommand{\Pexp}{{\cal P}\exp}
\newcommand{\Pa}{{\cal P}}

\newcommand{\bra}{\langle}
\newcommand{\ket}{\rangle}

\newcommand{\Tr}{{\rm Tr}}

\newcommand{\Or}{{\cal O}}

\newcommand{\ep}{\epsilon}

\newcommand{\al}{\alpha}

\newcommand{\la}{\lambda}

\newcommand{\ld}{\lambda_d}

\newcommand{\cur}{{\cal J}}

\begin{document}

\title{ Universal aspects of non-equilibrium currents in a quantum dot}

\author{Benjamin Doyon}
\affiliation{Rudolf Peierls Centre for Theoretical Physics, Oxford
University}
\author{Natan Andrei }
\affiliation{Center for Material Theory, Rutgers University}

\begin{abstract}

We study the electric current in the non-equilibrium Kondo model
at zero magnetic field, using real-time perturbation theory in the
Schwinger-Keldysh formulation. We show that the perturbative
coefficients to all orders have a finite limit at large switch-on
time ($t_0 \to -\infty$), and we give a prescription for general
operators to give finite coefficients in this limit. We explain
how this is related to the fact that the leads play the role of
thermal baths and allow relaxation to occur and the steady state
to form. This proves perturbatively that a steady state is reached
in the Schwinger-Keldysh formulation, and specifies which
operators correspond to quantities that have a well-defined value
in the steady state. Then, we show that the steady state can be
described by a special type of density matrix (related to
Hershfield's conjecture for the particular example of the
non-equilibrium Kondo model.) In the second part of the paper we
perform a renormalization-group analysis of the perturbative
series. We give a general argument that strongly suggests that the
perturbative series of any average in the steady state satisfies
the equilibrium Callan-Symanzik equations, and show in detail how
it works to one-loop order for the electric current operator
inside any average. We finally compute to two loops order the
average of the electric current in the steady state, and perform a
renormalization-group improvement. From this, we give a universal
prescription, valid in the perturbative regime, for comparing the
effect of the electric current to that of the temperature on the
``Kondo cloud''.

\end{abstract}

\maketitle

\bigskip

\sect{Introduction and discussion}

 The description of an out-of-equilibrium strongly correlated system
 is a long standing problem. Even in the simplest case where the
 system is in a steady state and its properties no longer change with
 time, the usual formalism of quantum statistical mechanics is
 inadequate. Theoretical understanding of such systems became all the
 more pressing with the recent spectacular progress in nanotechnology,
 which has made it possible to study the Kondo impurity, one of the
 best understood strongly correlated systems, in out-of-equilibrium
 conditions.

The Kondo impurity was realized experimentally as
 a quantum dot, a tiny island of electron liquid, attached via two
 tunnel junctions to leads (baths or reservoirs of electrons) held at
 different electric (or chemical) potentials. This set-up allows an
 electric current to flow across the dot, and measurements of the
 current were carried out as a function of the potential difference
 $V$, the temperature $T$ and the magnetic field $B$ \cite{exp}.

 When the dot carries a net spin in the Coulomb blockade regime, it
can be modeled by a Kondo Hamiltonian with two channels $\alpha =
1,2$, corresponding to the two leads, to which the  spin of the
dot, $\vec{S}$, couples \cite{qdot}. The resonant tunneling
through the dot (elastic co-tunneling) allows the electrons from
each bath to jump on the dot and back to the same bath, leading to
the formation of Kondo resonance around the Fermi level
$\mu_\alpha$ in each lead. Further, electrons from one bath can
jump on the dot and onto the other bath, giving ``off-diagonal
coupling'' of the two channels to each other. With the matrix of
couplings $J_{\alpha, \alpha'}$ and at zero magnetic field, the
Hamiltonian is
\beqa
H = &&\sum_{\alpha} \sum_{\vec{k},a} (\epsilon_{\vec{k}} -
\mu_{\alpha})c^{\dagger}_{\alpha, \vec{k},a}c_{\alpha,\vec{k},a} +
\sum_{\alpha, \alpha'}\sum_{\vec{k},\vec{k}', a ,a'}
     J_{\alpha, \alpha'}c^{\dagger}_{\alpha,\vec{k},a}{\bf \vec{\sigma}}_{a,a'}
     c_{\alpha',\vec{k}',a'} \cdot \vec{S}~.
\eeqa
Here $a$ denotes the spin index $a = \pm 1/2$, and $\vec{S}$ is in
the spin-1/2 representation.

The process corresponding to off-diagonal coupling induces a
current when the baths are held at nonzero potential difference,
$V=\mu_2 - \mu_1$. The development of the Kondo resonance as
temperature is lowered enables the system to overcome the Coulomb
blockade, producing a significant increase of the conductance. The
unitarity limit is reached as $T\to0,\,V\to0$.

As we are interested in the universal properties of the system, we
shall consider the model in the range $T, V \ll D_\alpha$, where
$D_\alpha=D$ are the bandwidths of the leads, each lead being
considered a very large conductor (the bandwidths can be assumed
to be the same for both channels). We are allowed therefore to
carry out the standard steps (linearizing around the Fermi level,
keeping only the $s$-wave component in the expansion of
$c_{\alpha,\vec{k},a}$ in spherical modes), to obtain a
representation of each lead as a free electron gas on the half
line consisting of left and right movers $\psi_{\alpha,L}(x),
\psi_{\alpha,R}(x), x \le 0$, interacting with the impurity
localized at $x=0$. It will be convenient for us to ``unfold'' the
baths, making left and right movers on the half line into right
movers on the full line (defining $\psi_{\alpha,R}(x)=
\psi_{\alpha,L}(-x), x \ge 0$). See Figure 1.

\begin{figure}[t]
\includegraphics[width=0.6\columnwidth, clip]{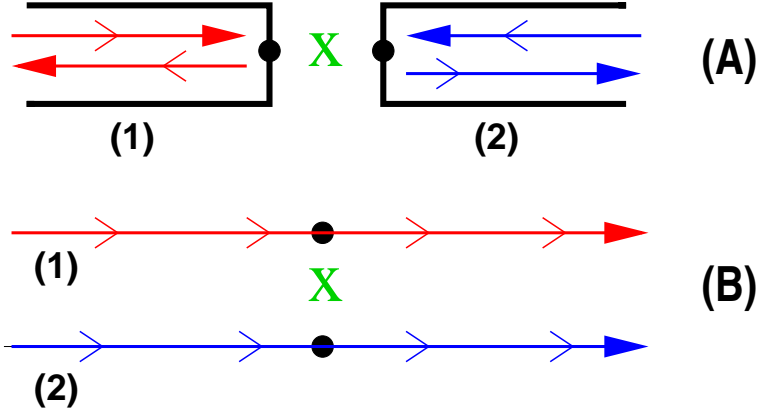}
\caption{ \label{fig:unfold} (Color online) (A) The system. (B) The system
unfolded.}
\end{figure}

The field-theoretic Hamiltonian is then:
\beqa
 H &=&
 -i \sum_{\alpha}\int_{-\infty}^{\infty} dx \,\psi^\dag_\alpha(x)\partial\psi_\al(x) +
\frc{V}2  \int_{-\infty}^{\infty} dx \;(\psi^\dag_2\psi_2 -
\psi^\dag_1\psi_1) + \sum_{\al,\al'=1,2} J_{\al,\al'}
\psi^\dag_\al(0) \vec{\sigma} \psi_{\al'}(0) \cdot \vec{S} \no \\
&=& H_0 + V\,H_z + H_I \label{Hfermions}
 \eeqa
where we work in units such that $v_F =1$. We also denote,
\[
H_z=\frc12\int_{-\infty}^{\infty} dx \;(\psi^\dag_2\psi_2 -
\psi^\dag_1\psi_1)
\]
 and
\[
H_I= \sum_{\al,\al'=1,2} J_{\al,\al'} \psi^\dag_\al(0)
\vec{\sigma} \psi_\al'(0) \cdot \vec{S}.
\]
The coupling of the baths to the quantum dot is parametrized by
the Hermitian matrix
 \beq\label{couplings}
J_{\al,\al'} = \pi\,\mato{cc} \ld & \la \\ \la & \ld
\matf_{\al,\al'}
\eeq
where $\ld$ and $\la$ are real (the factor $\pi$ is introduced for
later convenience). It is possible to diagonalize the matrix of
coupling $J_{\al,\al'}$ by a change of basis in the channel space.
In the simple situation of a single level quantum dot, described
by the Anderson model, coupled to identical leads one naturally
finds the relation $\ld = \la$. This makes the matrix of coupling
constants degenerate, one of the eigenvalues being zero. After
diagonalizing it, the terms representing the interaction with
$\vec{S}$ become that of a decoupled free fermion and a
one-channel Kondo model. More generally, for $\ld\neq\la$,
diagonalizing the matrix of couplings gives the interaction term
of the usual 2-channel Kondo model. Out of equilibrium, for $V\neq
0$, diagonalizing the matrix of couplings does not lead to a
simplification of the problem, since the out-of-equilibrium term
$V H_z$ induces extra coupling between the new fermion fields: it
is not invariant under change of basis. We will not perform this
diagonalization here, in order to keep the term $V H_z$ simple.
Also, we will consider the general case $\ld\neq\la$.

In this paper, we are interested in studying the electric current
as function of voltage and temperature in the steady state of this
model. As in the usual Schwinger-Keldysh formulation of a
non-equilibrium steady state \cite{Schwinger61,Keldysh65}, we will
imagine coupling the dot to the leads at some time $t_0$ in the
past when the system is in a thermal equilibrium state, then
allowing the system to evolve till $t=0$ when the current is
evaluated. One expects that after a transitory regime, as $t_0 \to
-\infty$, the system will relax into a steady state with a
constant current flowing through the dot.

Many questions arise concerning this formulation. Probably the
most obvious one is: Is the model sufficient to describe the
establishment of a steady state current? Or does one have to
contemplate additional relaxation mechanisms (certainly present in
actual experiments) to absorb the continuous flow of energy of the
electrons moving from the higher Fermi-level lead to the lower
one? In the framework of real-time
perturbation theory, a related (but not equivalent) question that
one can answer is whether or not the ``infrared'' limit
$t_0\to-\infty$ exists for the integrals representing the
perturbative coefficients in an expansion in $\la$ and $\ld$. We
develop the real-time Keldysh perturbation theory (in some ways similar to \cite{BazhanovLZ99}
and \cite{KaminskiNG00}), and use it to establish the convergence of every term of the perturbative
series as the switch-on time $t_0$ is sent to minus infinity.

This result is highly non-trivial. Real-time perturbation theory
often gives divergences as the switch-on time is sent to minus
infinity, unless a good relaxation phenomenon is included in the
model. In equilibrium, this issue can be easily overcome: an
infrared-divergent real-time perturbation theory for a system in
equilibrium only means that the particular model we are
considering does not have the proper relaxation mechanism. But
with an additional interaction, however small, representing a
proper relaxation mechanism, the system will have infrared
convergent real-time perturbation theory. Then, for equilibrium
models, this can be equivalently described by the
always-infrared-convergent ``imaginary time'' perturbation theory
(where the integrals in imaginary time are on a finite interval)
coming from the description of the model using its equilibrium
density matrix. There, the additional interaction representing
relaxation can be sent to zero from the beginning.

Out of equilibrium, however, there is {\em no a priori}
steady-state density matrix description of steady quantities.
Thus, if the real-time perturbation theory is infrared divergent,
there is no simple way to describe steady state physics. This is
very natural: in contrast to the equilibrium case, the relaxation
mechanism is used not only to {\em reach} the steady state, but
also to {\em form} it, since we need a continuous absorption of
energy. Hence, such infrared divergences are far more
pathological. If the limit of large negative switch-on time exists
order by order, this superficially seems a good indication that
the steady state is reached and that the model indeed describes
the steady state (although, strictly speaking, one would still have to analyse the possible non-perturbative
contributions if the perturbative series is asymptotic). But if no such limit exists, certainly more
questions arise: does the model reach a steady state
non-perturbatively (that is, divergences are an artifact of
perturbation theory), or are other interactions necessary? These questions of course
appeared in the literature before (see for instance \cite{WingreenS, rg6}).
An interesting example is the case studied in \cite{rg6}, where it was shown
that when the model we are discussing
is put into a magnetic field, the limit $t_0\to-\infty$ and $\lambda,\lambda_d\to0$
do not commute, leading to divergencies in the perturbation theory as $t_0\to-\infty$.
Assuming that a steady state exists in the model, the correct result as $t_0\to-\infty$
should then be non-perturbative, and it was partially evaluated under this assumption
(the ``zeroth order'' was evaluated), without external thermal reservoir coupled to the dot.

But the convergence of the perturbative series or the technical ``way around'' its divergencies
as just described above do not guarantee that the results
are describing the correct physical steady state. Indeed, if the model is believed to have a steady
state, perturbatively or not, two questions should still be answered: Is there 
an element in the model playing the role of a
good thermal reservoir to sustain the correct steady state in the model?
And if not, is there a guarantee that a coupling to an external reservoir would not have an important effect?

These last two questions have more bearing than it may
seem. In the usual Schwinger-Keldysh formulation, one {\em does not}
assume any exchange with a thermal bath while the system is
evolved: one starts with a thermal equilibrium state, then turns
on the coupling to the dot and lets the system evolve without
thermal bath. This is certainly not the true physical situation;
in fact, understanding how a thermal bath affects the evolution of
a quantum system was amongst the main points of the study of
Caldeira and Leggett \cite{CaldeiraL81}. We carry out the
real-time perturbation theory both in equilibrium (that is, in the
case $V=0$; note that the real-time formalism includes a
transitory non-equilibrium region where the system relaxes to
equilibrium) and out of equilibrium ($V\neq0$), and show that it
is the same phenomenon that makes the
perturbative series infrared convergent in equilibrium and out of
equilibrium. This phenomenon is a factorization
at large time separation of the correlation functions involved in
the perturbative coefficients; interpreting the integrals over time defining
these coefficients in terms of physical processes, this signals a decoherence in time
induced by the leads and suggests that the leads are {\em good thermal baths}.
The convergent expression in equilibrium is indeed the right equilibrium
density matrix, confirming that the leads themselves play the role of thermal baths
for the dot degrees of freedom. The convergent expression out of equilibrium
should then be ``the right'' steady state.
Note that the factorization signaling decoherence in time occurs
because the Hamiltonian for the leads is
conformally invariant: indeed, in conformal field theory, a large
time separation is a large distance separation, which, by locality, gives rise to factorization.
Physically, this occurs because the separation between the energy levels of the baths
is much smaller than all other scales in the problem and essentially energy-independent
(although, as will become clear from our investigation, these two conditions may not be sufficient).
Note that these ideas are not entirely new: in \cite{CallanT90} the results of Caldeira and
Leggett for constructing a thermal bath coupled to a quantum mechanical system
were re-interpreted as coupling a bulk conformal field theory in a disk to degrees of freedom on the boundary.
Our results generalize this to the behavior of the impurity in the Kondo model.
We are currently investigating how this can be further generalized (for example, what the general properties
of the conformal field theory should be).

Our proof also allows us to describe the steady-state physics in
terms of a ``steady state density matrix,'' as conjectured by
Hershfield \cite{Hershfield93}. The essential difference between
the usual density matrix and the steady-state density matrix can
be seen as a non-locality in the latter which captures
the build-up of the steady state.

Note that a proof of convergence to all order was developed in \cite{BazhanovLZ99} for a non-equilibrium
free boson model with
boundary interactions, but the arguments there were very model-dependent (and do not apply
to the non-equilibrium Kondo model) and quite different from ours.
In particular, our arguments have a much deeper physical meaning and scope, as explained above.

Let us stress here that it is quite important to know that no external bath is required
for reaching a non-equilibrium steady state in an impurity model. Indeed, this means that we can
use the real-time formalism {\em without addition of a coupling to an external bath} in order to
1) study the perturbative scaling properties of the model as we did in this
paper for the Kondo model (see below), 2) construct more or less
explicitly, without strong assumption, the steady state as an eigenstate of the Hamiltonian, for instance using the
the formalism of the ``steady-state density matrix''; this eventually can give access to the infrared
behavior of the model and to the integrability properties of the steady state (work in progress).
The real-time formalism might not be the easiest way of trying to obtain this understanding, but it is probably
the clearest, as it is the most closely related to the actual experimental situation.

Other questions that need to be addressed in the study of this
out-of-equilibrium steady state are: How will the Kondo effect,
the quenching of the impurity spin as the temperature is lowered
below the Kondo scale $T_K$, evolve in the presence of a current?
Will new scales make their appearance? Which quantities are
universal? To what extent the powerful ideas of the
Renormalization Group (RG) apply there? Many interesting attempts
were carried out, mainly perturbatively, to understand the flow of
couplings as the cut-off (band width) $D$ is reduced
\cite{rg1,rg2,rg3,rg4,rg5,rg6,rg7}.  We shall pursue a different
track and study a question related to the
universality features of the model, namely: does a limit $D \to
\infty$ exist? In this limit all results are universal. We
shall establish that such a limit exists by running the RG
equations ``backwards'', referring to them in the field theoretic
context, in the usual way, as the Callan-Symanzik equations. We
shall deduce an out-of-equilibrium $\beta$-function carrying out
the calculation directly in the steady state and will show that it
is the same as the equilibrium $\beta$-function. This may not be
too surprising since the singularity structure of the system
usually does not depend on the state in which they are evaluated,
so that the ground state and the highly excited steady state
produce the same singularities. The finite parts of course are
different. We show that only one scale arises, the Kondo
temperature, $T_K$, and the current can be written in a universal
form as a function of the ratios $ T/T_K, V/T_K, C$, with $C$  an
additional dimensionless parameter characterizing the asymmetry
between $\lambda$ and $\lambda_d$ (in other words, specifying the
RG trajectory). We carry out the computation of the current to
two-loop order and verify these statements explicitly.  We then
use the RG arguments to re-sum the leading logarithms.  Our
results are valid in the regime where both the bias voltage and
the temperature are smaller than the band width, and where the
bias voltage or the temperature is larger than the Kondo scale:
$V, T \ll D$, and $T_K\ll V$ or $T_K \ll T$. In particular, we
verify that there are no divergencies at $T/V\to0$ in the
perturbative results. This means that to two-loop order, the
voltage plays the role of a good infrared cutoff. From the RG
analysis, we give a universal prescription, valid in this regime,
for comparing the effect of the electric current to the effect of
the temperature on the destruction of the Kondo cloud.

We also examined the effect of a local magnetic field on the dot
but, as expected, were unable to show that the perturbation series converges in
this case. We will come back to a discussion of this case, in relation with the results
of \cite{rg6}, in the last section of this paper.

\section{Formulation of the problem and general considerations}

\label{sectKeldysh}

{\bf The Schwinger-Keldysh formulation
\cite{Schwinger61,Keldysh65}.}

{\em First formulation}.

We shall be interested in the
electric current  that passes from lead $2$ to lead $1$ across the
quantum dot under the action of the potential diffence $V$. It can
be calculated by evaluating the average of the current operator
$\cur$ with respect to a density matrix that has evolved over
sufficiently long time from the initial non-interacting density
matrix
\[
  \rho_0=   e^{-\beta H_0}
\]
under the action of the full evolution operator
\beq\label{evolop}
   S^{(V)}(t_1,t_2) = e^{-i(t_1-t_2) H}=e^{-i(t_1-t_2)( H_0 + V\,H_z + H_I)}~.
\eeq

The operational meaning of this formulation is the following. The
non-interacting leads are initially, say at time $t_0$, brought to thermal and
chemical equilibrium at zero potential difference exchanging
energy and particles with a common external reservoir at fixed
temperature and chemical potential. The energy levels of lead 1 and lead 2 are filled up to the same
energy (with thermal and particle fluctuations).

Just after time $t_0$, they are separated from
the external reservoir, then a potential difference $V$ is applied and
the interaction is turned on. The application of the potential $V$ just after time $t_0$, as usual, causes
a raising of the energy levels of lead 2 with respect to those of lead 1. For the clarity of the discussion below,
it is worth being more precise here. One should imagine both leads having
a continuum of available states from the bottom of their bandwidths with increasing energies (the energies
grow in a continuous way for infinite leads, of course, so one should think about densities of states). At time
$t_0$, the available states of the leads are filled up to equal energies.
Then, just after time $t_0$, when the potential is applied, one shifts the energies of all states of lead 2
by, say, $V/2$ (towards higher energies), and
the energies of those of lead 1 by $-V/2$, without changing the occupations of the states. Hence,
the levels of lead 2 are now filled up to a higher energy than those of lead 1. Since the reservoir is disconnected
and the interaction is turned on,
there is a current. The steady state current is obtained after an infinite time, which we will
take to be time 0 (that is, we will take $t_0\to-\infty$).

In the equation (\ref{evolop}), the raising of the energy levels and the turning on of the
interaction strength seem instantaneous and simultaneous. But one can multiply both terms $VH_z$ and $H_I$
by a factor that smoothly increases from 0 at time $t_0=-\infty$ to 1 at time $0$, for instance the factor
$e^{-\mu t}$,
in order to implement a simultaneous adiabatic increase of both the potential and the interaction strength.
Sending $\mu\to 0$ (the adiabatic increase occurring far in the past) gives the steady state.
This is really what is understood in this formulation.

It is not obvious, a priori, that this formulation represents
adequately the usual experimental situation, where the leads and
the quantum dot are always connected to a common thermal reservoir
(but not a reservoir of electrons), even while the steady state is
being reached. However, it is natural to think that the leads can
themselves play the role of thermal reservoirs. As discussed in
the introduction, this is indeed the case, and will be made more
precise below.

The average of an operator $\Or$ in the steady state is then given by
\beq\label{opss}
    \bra \Or \ket_{ss} =
    \lim_{t_0 \to-\infty} \frc{\Tr\lt[
        S^{(V)}(0,t_0)\;e^{-\beta H_0}\;S^{(V)}(t_0,0)\; \Or
        \rt]}{\Tr\lt[ e^{-\beta H_0} \rt]}~.
\eeq

The operators act in the Hilbert space for $H_0$ (which is a
tensor product of the two-channel free massless fermion Hilbert
space and of the impurity space) obtained by imposing
asymptotically vanishing conditions for the fermion fields
(correlation functions of fermion fields vanish at infinite
distance from each other and from the dot). To be more accurate,
we could start by taking the fermion fields on a line segment of
length $L$ containing the dot, with some free boundary conditions;
for instance, $\psi_\alpha(-L)=\psi_\alpha(L)$ (this corresponds
to the usual free boundary conditions when fermions are folded
back on the half line), then send $L$ to infinity. The steady
state would be obtained in the region
\beq\label{orderlimits}
    L^{-1} \;\ll\; |t_0|^{-1} \;\ll\; V, T
\eeq
where the energy scale of switch-on, $|t_0|^{-1}$, suffices to
smear out the energy level spacing $L^{-1}$.

{\em Second formulation}.

Another formulation can be given. The initial idea of this second formulation is that the current can be created not
only by a shift of the energies of the states of leads 1 and 2 (coming from the application
of an electric potential), but also by {\em putting additional
electrons in lead 2 and taking away electrons from lead 1}. In order to implement this,
one starts again, at time $t_0$, with the uninteracting
leads, both connected to a common thermal and particle reservoir, and in themal and
chemical equilibrium; but now the chemical equilibrium is not at potential difference 0, but rather
at a potential difference $-V$. The initial density matrix is then
\[
    \tilde{\rho}_0= e^{-\beta(H_0-VH_z)}~.
\]
This potential difference shifts towards lower energies the states of
lead 2 with respect to those of lead 1 by an amount $V$. But since there is equilibrium,
the states of lead 1 and lead 2 are still filled up to the same energy. Note that then, as compared to the
first formulation at time $t_0$, there are more available sates of lead 2 and less of lead 1 that are filled.

Just after time $t_0$, the reservoirs are disconnected, then the potential is set to 0 and the interaction
is turned on. The density matrix $\tilde{\rho}_0$
then evolves with the evolution operator at zero bias voltage $\bar{S}(t_1,t_2)$,
\beq\label{S}
    \bar{S}(t_1,t_2) = S^{(V=0)}(t_1,t_2)=   e^{-i(t_1-t_2)( H_0  + H_I)}  ~.
\eeq
Putting the potential to 0 has the effect of raising the energy levels of lead 2 with
respect to those of lead 1 by an amount $V$, the same effect that occurs in the first formulation just after
time $t_0$ when the potential is applied. In contrast, though, this brings
us to a situation where the available states of leads 1 and 2 have exactly the same energies
as in the first formulation {\em at time $t_0$} (that is, at potential 0),
but with {\em more
states filled in lead 2 and less in lead 1}, so that the leads are filled up to unequal energies.
This indeed implements having put additional electrons in lead 2
and extracted electrons from lead 1. With the interaction on and the reservoir disconnected,
a current is created. Again, after an infinite time, the steady state should be reached. The current is then given by
\beq\label{curssV}
    \bra \cur \ket_{ss} =
    \lim_{t_0 \to-\infty} \frc{\Tr\lt[
       \bar{ S}(0,t_0)\;e^{-\beta (H_0-VH_z)}\;\bar{S}(t_0,0)\; \cur
        \rt]}{\Tr\lt[ e^{-\beta (H_0-VH_z)} \rt]}
\eeq
(for more general operators $\Or$, see (\ref{opssV})).

If the size of the bandwidth can be sent to infinity (when evaluating quantum averages of operators that give
finite results in this limit), then the operational description above for the second formulation is equivalent to
that of the first formulation, since then only the
Fermi energies of leads 1 and 2 matter. In particular, raising the energy levels or filling states with electrons are
exactly the same operation in this case.

However, there is another difference between both formulations.
In the second formulation, we can now think about putting a factor
$e^{-\mu t}$ for adiabatically increasing the interaction strength, but there is no such possibility
for adiabatically increasing the potential (one would have to add the term $-(1-e^{-mu t})VH_z$ in the evolution
Hamiltonian). In other words, for practical calculations,
this second formulation naturally implies that the energy levels of lead 2 are raised (or the states are filled),
with respect to those of lead 1,
by an amount $V$ instantaneously, and that the interaction is then turned on adiabatically. This is to be
contrasted with the first formulation, where both the potential difference and the interaction strength
were understood as being simultaneously increased adiabatically.

We will show below that both formulations are equivalent.

\bigskip

{\bf Symmetry currents.} The Hamiltonian $H_0$ is conformally
invariant and has a large algebra of symmetries associated with
it. It is a WZW ``current algebra'' of the symmetry currents (not
to be confused with the physical current $\cur$) and it will be
convenient to carry out many of the calculations in terms of
symmetry currents. Introduce the following operators,
\beqa
    J_z &=& \pi :(\psi^\dag_2\psi_2 - \psi^\dag_1\psi_1):
    \no\\
    \vec{J}_x &=& i\pi (\psi^\dag_2\vec{\sigma}\psi_1 -
    \psi^\dag_1\vec{\sigma}\psi_2) \no\\
    \vec{J}_y &=& \pi (\psi^\dag_2\vec{\sigma}\psi_1 +
    \psi^\dag_1\vec{\sigma}\psi_2) \no\\
    \vec{J}_d &=& \pi :(\psi^\dag_2\vec{\sigma}\psi_2 +
    \psi^\dag_1\vec{\sigma}\psi_1): ~.
\eeqa
They form the following subalgebra of the $su(4)_1$ current
algebra:
\beqa\label{algxyz}
    [J_d^i(x),J_d^j(y)] &=& 2i\pi \lt(\epsilon_{ijk}\; J_d^k(x)
    \;\delta(x-y) -  \delta_{ij} \;\delta'(x-y)\rt)
    \no\\ {}
    [J_x^i(x),J_x^j(y)] &=& 2i\pi \lt( \epsilon_{ijk} \; J^k_d(x)
    \;\delta(x-y) - \delta_{ij} \;\delta'(x-y)\rt)
    \no\\ {}
    [J_y^i(x),J_y^j(y)] &=& 2i\pi \lt(\epsilon_{ijk} \; J^k_d(x)
    \;\delta(x-y) - \delta_{ij} \;\delta'(x-y)\rt)
    \no\\ {}
    [J_z(x),J_z(y)] &=&  - 2i\pi \;\delta'(x-y)
    \no\\ {}
    [J_d^i(x),J_x^j(y)] &=& 2i\pi \;\epsilon_{ijk} \; J^k_x(x) \;\delta(x-y)
    \\ {}
    [J_d^i(x),J_y^j(y)] &=& 2i\pi \;\epsilon_{ijk} \; J^k_y(x) \;\delta(x-y)
    \no\\ {}
    [J_d^i(x),J_z(y)] &=& 0
    \no\\ {}
    [J_x^i(x),J_y^j(y)] &=& 2i\pi \;\delta_{ij} \; J_z(x) \;\delta(x-y)
    \no\\ {}
    [J_y^i(x),J_z(y)] &=& 2i\pi \;J_x^i(x) \;\delta(x-y)
    \no\\ {}
    [J_z(x),J_x^i(y)] &=& 2i\pi \;J_y^i(x) \;\delta(x-y)~.
    \no {}
\eeqa
 In terms of these currents
the full Hamiltonian $    H = H_0 + V\,H_z + H_1$ can be expressed
as follows:
\beqa
\displaystyle    H_0 &=& \frc1{4\pi} \int_{-\infty}^\infty
dx\,\lt(
        :\vec{J}_d(x)\cdot \vec{J}_d(x):
    + :\vec{J}_x(x)\cdot \vec{J}_x(x): + :\vec{J}_y(x)\cdot \vec{J}_y(x):
    + :J_z(x)\cdot J_z(x):\rt)~, \n
\displaystyle    H_z &=& \frc1{2\pi} \int_{-\infty}^{\infty}dx\,
    J_z(x) \n
\displaystyle    H_I &=&
    \lambda_d \vec{J}_d(0)\cdot \vec{S} +
    \lambda \vec{J}_y(0)\cdot \vec{S}~.
\eeqa
The operator $H_z$ is the total (normalized) isospin
$z$-component. \footnote{One could also introduce the operator
$J_c = \pi(\psi^\dag_2\psi_2 + \psi^\dag_2\psi_1)$. Since the
total (sum over the two channels) charge is locally conserved, the
current $J_c$ commutes with all current operators introduced
above, hence it decouples from the theory and can be identically
set to 0.}

{\bf The electric current.} We now turn to discuss in more detail
the current across the quantum dot (we set the electric charge
$e=1$), and express it also in terms of the symmetry currents. The
electric current $\cur$ is given by  the operator measuring the
difference between the fermion density on, say, the second channel
just before hitting the impurity and the fermion density on the
same channel just after hitting it:
 \beq\label{current}
\cur = \lim_{\epsilon\to0^+} \lt(\psi^\dag_2\psi_2(x=-\epsilon) -
\psi^\dag_2\psi_2(x=\epsilon) \rt) = \frc1{2\pi}
\lim_{\epsilon\to0^+} (J_z(-\epsilon) - J_z(\epsilon))~.
 \eeq
(we use here the ``unfolded set-up''). Equivalently one can
express the electric current as the rate of decrease of the charge
on lead-2 (or increase on lead-1),
\beqa\label{curHz}
\cur &=& - \frc{d}{dt}N_2 = \frc{d}{dt}N_1 = - i[H,H_z]\nonumber\\
   &=&\lambda \vec{J}_x(0)\cdot \vec{S}~.
\eeqa

It is easy to see that the two definitions coincide. We rewrite
the first definition using ``impurity conditions'' - operator
relation inherited from boundary conditions. Boundary conditions,
in general, are part of the equations of motion and lead to
operator relations valid on the full Hilbert space of a boundary
quantum field theory. They are often derived from the action of
the model in the same way as one derives the equations of motion.
From our view point, after ``unfolding'' the Kondo model, we have
a model with an impurity instead of a boundary. As with
boundaries, impurities give rise to ``impurity conditions'' which
are part of the equations of motion and are operator relations
valid on the full Hilbert space. In operator language (which is
more convenient for our purposes), the impurity condition
associated to a local operator $\Or(x)$ can be written
\[
    \lim_{\epsilon\to 0^+} \lt(\int_{-\infty}^{-\epsilon}dx +
    \int_{\epsilon}^{\infty}dx\rt)\, [H,\Or(x)] =
    \int_{-\infty}^\infty[H,\Or(x)]~.
\]
Consider the impurity condition associated to the operator
$J_z(x)$ with the model with Hamiltonian $H$ (\ref{Hfermions}),
\beq\label{impurcond}
    \lim_{\epsilon\to 0^+} \lt(\int_{-\infty}^{-\epsilon}dx +
    \int_{\epsilon}^{\infty}dx\rt)\, [H,J_z(x)] =
    \int_{-\infty}^\infty[H,J_z(x)]~.
\eeq
On the left-hand side, only the free part $H_0$ of the Hamiltonian
is involved, because the operator $J_z(x)$ is never at the site
$x=0$ (and $H_z$ commutes with $J_z(x)$). Using the fact that with
the free Hamiltonian $H_0$, $J_z(x)$ is a right-moving operator
$[H_0,J_z(x)] = i\frc{d}{dx}J_z(x)$, and using the asymptotic
conditions $J_z(\infty) = J_z(-\infty)$, we have
\[
    \lim_{\epsilon\to 0^+} \lt(\int_{-\infty}^{-\epsilon}dx +
    \int_{\epsilon}^{\infty}dx\rt)\, [H,J_z(x)] =
    \lim_{\epsilon\to 0^+} i(J_z(-\epsilon) - J_z(\epsilon))~.
\]
On the other hand, on the right-hand side of (\ref{impurcond}),
since the integration is on the full interval, the free part of
the Hamiltonian does not contribute. Only the impurity term, at
$x=0$, contributes, and it gives
\[
    \int_{-\infty}^\infty[H,J_z(x)] =
    2i\pi \lambda \vec{J}_x(0)\cdot \vec{S}~.
\]
as expected.

Having now discussed the system and the various operators
describing it  we turn to discuss in more detail the nature of
non-equilibrium in the system.

\section{Equilibrium vs. non-equilibrium}

Our model, a quantum impurity coupled to leads at different
chemical potentials, describes a non-equilibrium situation -- a
current is flowing from one lead to another. What is the the
precise meaning of this statement?

 In this section we show in what sense an out-of-equilibrium
model differs from an equilibrium model.  We begin by showing how
the Keldysh formulation  leads, when the system is in equilibrium,
to the usual equilibrium density matrix description. We prove, in
other words, the following:
\beq\label{equil}
    \frc{\bar{S}(0,-\infty) e^{-\beta H_0}
    \bar{S}(-\infty,0)}{\Tr \lt[e^{-\beta H_0}\rt]}  =
    \frc{e^{-\beta H_0}\Pexp\lt(i\int_{-i\beta}^{0}dt\,
    H_I^{(0)}(t) \rt)}{\Tr\lt[e^{-\beta H_0}\Pexp\lt(i\int_{-i\beta}^{0}dt\,
    H_I^{(0)}(t) \rt)\rt]}  = \frc{e^{-\beta H|_{V=0}}}{\Tr\lt[e^{-\beta
    H_{V=0}}\rt]}
\eeq
as an equation to hold when evaluated inside traces with insertion of any number of
local operators at fixed positions.

A local operator is, by definition, an operator
depending on the position $x$ (in the sense that its commutator with the momentum
operator is a derivative with respect to $x$), such that its commutator with the hamiltonian density
at position $y$ is zero for $x\neq y$. Note that local charges, for instance conserved
charges of the Hamiltonian, are integrals of local operators, and are not local operators
themselves. Hence, the limit (\ref{equil}) does not hold with insertion of local charges.
This makes physical sense, since conserved charges are not expected to relax to their equilibrium
values. Technically, one must remember that the density matrix
is an operator with infinitely many matrix elements, hence any limit applied to it cannot be expected to converge
to an object having the same properties (or to converge at all)
independently from which subset of matrix elements we are looking at.

This derivation is important for what follows, so we  present it
in some detail. Recall that $\bar{S}(t_1,t_2)$, Eq. (\ref{S}), is
the evolution operator at zero voltage. We now establish some
useful identities.  In the interaction picture with respect to
$H_0$ we have,
\beqa
 \bar{S}(0,t_0) e^{-iH_0t_0}e^{-\beta H_0} &=&
\Pexp\lt(i\int_{0}^{t_0}dt \, H_I^{(0)}(t) \rt) e^{-\beta H_0} =
 e^{-\beta H_0} \Pexp\lt(i\int_{-i\beta}^{t_0-i\beta}dt \,
H_I^{(0)}(t) \rt) \label{aaa}
\eeqa
where in the interaction picture
\[
    H_I^{(0)}(t) = e^{iH_0 t} H_I e^{-iH_0 t}
     = \lambda_d\vec{J}_d(-t)\cdot \vec{S}
        + \lambda \vec{J}_y(-t) \cdot \vec{S} ~.
\]
In (\ref{equil}) and in the last two expressions of (\ref{aaa}),
the symbol $\Pa$ indicates path-ordering in time: the operators
are positioned from left to right with their time argument going
from the lower integral limit to the upper integral limit. In the
first occurrence in (\ref{aaa}), integrals are ordered from $0$ on
the left to $t_0$ on the right. In the second, the integration
contour is from $-i\beta$ on the left to
$t_0-i\beta$ on the right. On the other hand, we have,
\beq\label{bbb}
    e^{iH_0 t_0} \bar{S}(t_0,0) =
    \Pexp\lt(i\int_{t_0}^{0}dt\,
    H_I^{(0)}(t) \rt)~.
\eeq

The Keldysh evolution is then (multiplying (\ref{bbb}) with
(\ref{aaa}) and dividing by the trace of this product),
\beqa \frc{\bar{S}(0,t_0) e^{-\beta H_0} \bar{S}(t_0,0)}{\Tr\lt(e^{-\beta
H_0}\rt)} &=& \frc{e^{-\beta H_0}
\Pexp\lt(i\int_{-i\beta}^{t_0-i\beta}dt \, H_I^{(0)}(t) \rt)\
\Pexp\lt(i\int_{t_0}^{0}dt\, H_I^{(0)}(t) \rt)}{ \Tr\lt[e^{-\beta
H_0} \Pexp\lt(i\int_{-i\beta}^{t_0-i\beta}dt \, H_I^{(0)}(t) \rt)\
\Pexp\lt(i\int_{t_0}^{0}dt\, H_I^{(0)}(t) \rt) \rt]} \n
&\stackrel{|t_0|\gg\beta}=& \frc{e^{-\beta H_0}
\Pexp\lt(i\int_{-i\beta}^{t_0}dt \, H_I^{(0)}(t) \rt)\
\Pexp\lt(i\int_{t_0}^{0}dt\, H_I^{(0)}(t) \rt)}{ \Tr\lt[e^{-\beta
H_0} \Pexp\lt(i\int_{-i\beta}^{t_0}dt \, H_I^{(0)}(t) \rt)\
\Pexp\lt(i\int_{t_0}^{0}dt\, H_I^{(0)}(t) \rt) \rt]}~.
\label{evolveddm}
\eeqa
The last equality is valid perturbatively if the integration from
$t_0-i\beta$ to $t_0$ is negligible at every order in perturbation
theory.

Note that the last equality involves taking $|t_0|$ much greater than $\beta$.
At zero temperature, when $\beta\to\infty$, this condition cannot hold, and since the correlation
functions then may have algebraic decay with power $-1$ at large distances, our proofs below (at equilibrium and
in the steady state) do not apply. Nevertheless, as will be seen, our two-loop perturbative results
for the non-equilibrium current have finite zero-temperature limit; this will be discussed further in the last
section.

To show the last equality in (\ref{evolveddm}) we evaluate the expectation value of a local
operator (or product of any local operators at fixed positions)
$\Or$, inserted at the right-hand side of the first equation of
(\ref{evolveddm}). Denoting by
\beq\label{freetrace}
\bra\bra\cdots\ket\ket_0 = \frc{\Tr\lt(e^{-\beta
H_0}\cdots\rt)}{\Tr\lt(e^{-\beta H_0}\rt)}
\eeq
the averaging in the free theory at temperature $\beta^{-1}$, we
consider,
\beq\label{ddd}
\frc{\lt\bra\lt\bra\Pexp\lt(i\int_{-i\beta}^{t_0-i\beta}dt \,
H_I^{(0)}(t) \rt)\ \Pexp\lt(i\int_{t_0}^{0}dt\, H_I^{(0)}(t) \rt)\
\Or \rt\ket\rt\ket_0}{ \lt\bra\lt\bra
\Pexp\lt(i\int_{-i\beta}^{t_0-i\beta}dt \, H_I^{(0)}(t) \rt)\
\Pexp\lt(i\int_{t_0}^{0}dt\, H_I^{(0)}(t) \rt) \rt\ket\rt\ket_0}~.
\eeq
All correlation functions  involved are correlation functions
where the $H_I(t)$'s are {\em connected} to $\Or$. Connected
correlation functions are defined, in the usual way, by
subtracting from correlation function appropriate products of
expectation values. In Appendix \ref{connected} we recall their precise
definition and main properties. Only connected correlation
functions occur, because in (\ref{ddd}) we divide by the
correlation function of the operator where all the $H_I(t)$'s are
involved.

In Appendix \ref{factorization} we show that correlation functions of the type
\[
    \bra\bra H_I^{(0)}(t+t_1) H_I^{(0)}(t + t_2) \cdots H_I^{(0)}(t+t_n)
    \Or\ket\ket_0
\]
factorize, as $t\to\pm\infty$, into
\[
    \bra\bra H_I^{(0)}(t_1) H_I^{(0)}(t_2) \cdots H_I^{(0)}(t_n)\ket\ket_0\;
    \bra\bra\Or\ket\ket_0~,
\]
with sub-leading asymptotic contributions vanishing exponentially
for finite $\beta$.

The last step of (\ref{evolveddm}) follows from the factorization
property. Due to this property, connected correlation functions of
the type
\[
    \bra\bra H_I^{(0)}(t_1) H_I^{(0)}(t_2) \cdots H_I^{(0)}(t_n) \Or\ket\ket_{0,\rm connected}
\]
vanish exponentially whenever any subset of consecutive time
variables $\{t_i,t_{i+1},\ldots,t_j\}$ (corresponding to a subset
of time-ordered operators $H_I(t)$'s) goes to negative infinity
simultaneously. This implies that order by order in perturbation
theory of (\ref{ddd}), all integrands vanish exponentially in any
large-time region, in particular in the segment $t_0 -i\beta$,
which then factorizes and cancels between numerator and
denominator.  Hence, in the limit where $t_0 \to -\infty$ the last
step of (\ref{evolveddm}) is exact order by order in perturbation
theory, and we have (\ref{equil}), as claimed. See Figure 2.

\begin{figure}[t]
\includegraphics[width=0.6\columnwidth, clip]{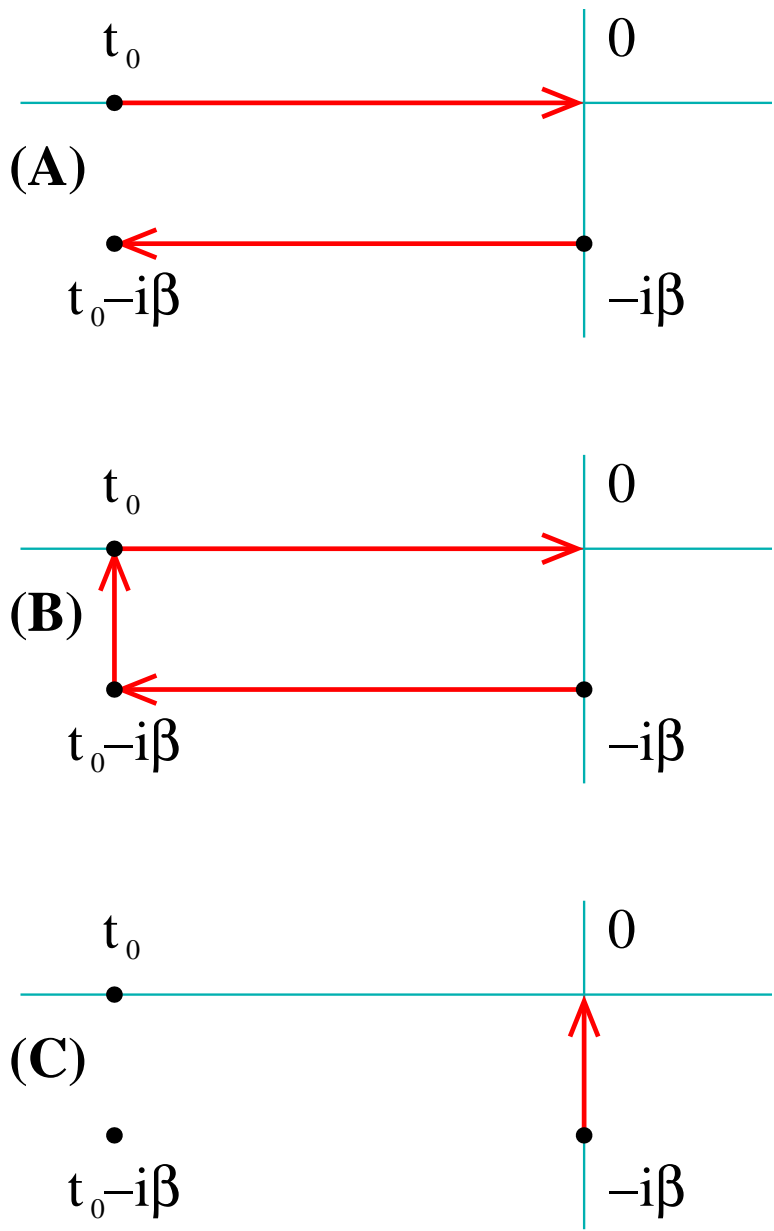}
\caption{ \label{fig:noneqtoeq}
(Color online) Deriving the equilibrium formulation starting from real time
(Keldysh) formulation.  (A) The Keldysh integration contour on the complex time plane;
the directions of the arrows indicate the ordering of operators from left to right. (B)
Adding the segment $(t_0, t_0 -i \beta)$ is allowed by the
factorization property (this step is not allowed when the system
is out of equilibrium). (C) The equilibrium contour.}
\end{figure}

We wish to note that our argument in Appendix  \ref{factorization}  relied on the
fact that $H_I(t)$ is a local, right-moving operator, and that it
couples to the external degree of freedom $\vec{S}$ (the impurity)
in an $SU(2)$-invariant way. For electronic degrees of freedom
 large time means large distance, and at large
distances, correlation functions of local fields factorize.
Combined with $SU(2)$ invariance, this implies the factorization
of correlation functions of $H_I(t)$'s at large $|t|$.

The implications of the well-defined limit $t_0/\beta\to-\infty$,
and in particular of the factorization at large time separation of
the correlation functions involved in the perturbative
coefficients, are important. It was not necessary to invoke any
external relaxation mechanism: the factorization signals a
decoherence in time and suggests that $H_0$ represents a good
thermal bath, and this bath by itself provides such a mechanism.
As in Caldeira-Leggett models, $H_0$ can be seen as an infinity of
free oscillators with an appropriate frequency distribution in
order to represent a thermal bath. The loss of time-reversal
symmetry associated with this relaxation mechanism occurs when
taking the limit $|t_0|/\beta\to\infty$.

The same derivation can be carried out for more general unitary
conformal field theories perturbed by an interaction $H_I$ at one
point, or  defined on a finite region of space. Inferring from our derivation, the interaction
can be due to an external degree of freedom coupled to any linear
combination of fields that factorize into right- and left-movers,
and the coupling has to be invariant with respect to a symmetry
group acting on the full configuration space.

{\bf The steady state current.} The derivation fails when out of
 equilibrium, $V\neq0$.  The step that becomes incorrect, if we start with expression (\ref{opss})
for the steady-state average, is the
 shifting of the integration limits $t_0-i\beta \to t_0$. Indeed, the
 correlation functions involving $V H_z^{(0)}(t)=V H_z$ are not
 suppressed at large negative times since $H_z$ is a conserved charge
 of the Hamiltonian $H_0$. Moreover, due to quantum fluctuations of
 the charge $H_z$, made possible by the interaction $H_I$ (that is,
 $[H_I,H_z]\neq0$), connected correlation functions involving $H_z$
 are not zero. These two conditions are at the origin of the
 appearance of a non-equilibrium situation. In physical terms, the
 first condition is that the bath represented by $H_0$ does not
 provide a relaxation mechanism for reaching Boltzmann's distribution
 of states associated to the energy $H_0+VH_z+H_I$; the second
 condition is that nevertheless, $H_z$ is subject to quantum
 fluctuations and evolves with time. In this case then the Keldysh
 formulation does not reduce to an equilibrium description.

Our analysis, however, has not yet established that a steady state
occurs. We shall present below a full proof to this effect. To
motivate the proof we begin with a physical argument, based on the first formulation
described around Eqs. (\ref{evolop}) and (\ref{opss}), by
considering the respective ground states of $H_0$ and $H$ (instead
of the associated thermal density matrices) and showing that they
are ``far'' enough and that the evolution of $H_z$ is ``slow''
enough in the limit $L\to \infty$ so that a steady state is
established.  Under other circumstances, we might expect some
oscillating behavior.

That the ground state of $H$ is far enough, and the evolution of
$H_z$ is slow enough, can be made more precise in the following
way. To begin with, consider the ground state $|0\ket$ of $H_0$
and the ground state $|V\ket$ of the Hamiltonian   $H_0 +V  H_z$.
Later we shall consider the effect of the couplings $\lambda$ and
$\lambda_d$. The ground state $|V\ket$ can be obtained in the
following way. Consider the operator
\beq\label{trans}
    U_{V} = e^{i\frc{V}\pi\int_{-\infty}^{\infty} dx\,xJ_z(x)}
\eeq
(in order for it to be well defined, we assume an appropriate
ultraviolet regularization of the operator $J_z(x)$). This is a
unitary operator, and its effect on $H_0$ is:
\beqa
    U_{-V} e^{-\beta H_0} U_{V} &=&e^{-i\frc{V}\pi \int dx\, xJ_z(x)}
e^{-\beta H_0}e^{i\frc{V}\pi \int dx\, xJ_z(x)}\no \\
&=&e^{-\beta H_0} e^{-i\frc{V}\pi\int
    dx\,x J_z(x+i\beta)} e^{i\frc{V}\pi \int dx\, xJ_z(x)} \no\\
    &=& e^{-\beta H_0} e^{-i\frc{V}\pi\int
    dx\,(x-i\beta) J_z(x)} e^{i\frc{V}\pi \int dx\, xJ_z(x)} \no\\
    &=& e^{-\beta H_0} e^{-\beta VH_z} e^{\frc{V^2}{2\pi^2}
    \int dx_1dx_2\, (x_1-i\beta)x_2\,[J_z(x_1),J_z(x_2)]}~.
    \label{transH0}
\eeqa The last exponential factor is a real number, scaling with the
system size $L$ (it could be absorbed into the definition of
$U_{-V}$, at the price of losing its unitarity). Hence, the
operator $U_{-V}$ takes $H_0$ to $H_0+VH_z$, up to an additive
number, and the ground state of $H_0+VH_z$ can be obtained by
\beq
    |V\ket = U_{-V} |0\ket~.
\eeq

Computing the expectation values of $H_z$ (tracing over the
two-dimensional impurity space) in these ground states we have, as
$L\to\infty$,
\beq
    \bra 0|H_z|0\ket = 0~,\qquad \bra V|H_z|V \ket \sim -VL~.
\eeq
 In particular, $U_{-V}$ has the effect, in the infinite-$L$ limit, of
changing the asymptotic conditions to $J_z(\pm\infty)=-\pi V/2$.
We discussed the ground state of $H_0 +V H_z$. However, the actual
ground state of $H$ yields corrections to the expectation values
that are of higher order in the couplings with a finite limit as
$L\to\infty$, and our conclusion therefore apply to the full
Hamiltonian.

While the expectation values are infinitely distant {\em in an
infinite system}, the rate of change of $ \bra H_z\ket $ as the
interaction is switched on is finite since the operator $H_I$
giving rise to the current is local. This will be seen explicitly
in the perturbative calculations of the current $\cur$ below
(recall that $\cur = - i[H,H_z]$).

Hence, as $L\to\infty$, it would take more and more time to get
from $|0\ket$ to $|V\ket$. Here we assume that the average of
$H_z$ would decrease monotonically. This is expected for $L$ large
enough and elapsed time large enough, though not infinite. More
precisely in the region (\ref{orderlimits}), we expect the
expectation varlue of $H_z$ to decrease steadily; this is the
steady state. In other words, we expect a steady state to occur
because $H_z$ scales with the length of the system, whereas its
variation does not.  For a finite $L$, it does not decrease
monotonically at all times, and we might eventually see an
oscillating behavior of period characterized by $L$.

We proceed now to the main result of this section: we show that to
all orders in perturbation theory the limit of very large negative
times, $t_0/\beta\to-\infty$ in (\ref{opss}), is well defined for
any local operator, $\Or$,  supported at a point or  on a finite
interval. This shows that there is indeed a steady state: the
current operator $\cur$ acquires a well-defined expectation value.

Using the interaction picture with respect to $H_0 + VH_z$, we can
write the steady-state average of any operator $\Or$ as
\beq\label{ssV}
    \bra\Or\ket_{ss} = \lim_{t_0/\beta\to-\infty}
        \frc{1}{\Tr\lt[e^{-\beta H_0}\rt]}\Tr\lt[
        \Pexp\lt(i\int_{0}^{t_0}\hspace{-3mm}dt\,
        H_I^{(V)}(t)\rt)\,
         e^{-\beta H_0}\,
        \Pexp\lt(i\int_{t_0}^0\hspace{-3mm}dt\, H_I^{(V)}(t) \rt) \;
        \Or
        \rt]
\eeq
with
\beq\label{H1Vt}
    H_I^{(V)}(t) = e^{i\lt(H_0 + V H_z\rt)t} \;H_I \;
    e^{-i\lt(H_0 + V H_z\rt)t} =
        e^{iVH_zt}H_I^{(0)}(t)e^{-iVH_z t}~.
\eeq
The operator $H_I^{(V)}(t)$ can be expressed in terms of
``deformed'' current-algebra operators. Consider
\beqa
    \vec{J}_d^{(V)}(x) &=& e^{-iVH_zx}\vec{J}_d(x)e^{iVH_zx}\n
    \vec{J}_x^{(V)}(x) &=& e^{-iVH_zx}\vec{J}_x(x)e^{iVH_zx}\n
    \vec{J}_y^{(V)}(x) &=& e^{-iVH_zx}\vec{J}_y(x)e^{iVH_zx}~.\no
\eeqa
It is a simple matter to use the commutation relations
(\ref{algxyz}) in order to obtain
\beqa
    \vec{J}_d^{(V)}(x) &=& \vec{J}_d(x) \n
    \vec{J}_x^{(V)}(x) &=& \cos(Vx) \vec{J}_x(x) + \sin(Vx)
        \vec{J}_y(x) \n
    \vec{J}_y^{(V)}(x) &=& -\sin(Vx) \vec{J}_x(x) + \cos(Vx)
        \vec{J}_y(x)~.
        \label{JV}
\eeqa
Using these operators, we have
\beq
    H_I^{(V)}(t) = \lambda_d\vec{J}_d^{(V)}(-t)\cdot \vec{S} +
    \lambda \vec{J}_y^{(V)}(-t) \cdot \vec{S}~.
\eeq

The proof that the limit $t_0/\beta\to-\infty$ exists in
(\ref{opss}) proceeds from arguments similar to those in the
previous subsection. Let the operator $\Or$ be supported on a
finite interval. Moving the operator $e^{-\beta H_0}$  to the left
inside the trace in (\ref{ssV}), we have
\beq
    \bra\Or\ket_{ss} = \lim_{t_0/\beta\to-\infty}
        \frc{\Tr\lt[
        e^{-\beta H_0}\,\Pexp\lt(i\int_{-i\beta}^{t_0-i\beta}dt\,
        \tilde{H}_1^{(V)}(t)\rt)\,
        \Pexp\lt(i\int_{t_0}^0dt\, H_I^{(V)}(t) \rt) \;
        \Or
        \rt]}{\Tr\lt[
        e^{-\beta H_0}\,\Pexp\lt(i\int_{-i\beta}^{t_0-i\beta}dt\,
        \tilde{H}_I^{(V)}(t)\rt)\,
        \Pexp\lt(i\int_{t_0}^0dt\, H_I^{(V)}(t) \rt) \;
        \rt]}
\eeq
where
\beqa
    \tilde{H}_I^{(V)}(t) &=& e^{-\beta V H_z}H_I^{(V)}(t) e^{\beta
    VH_z} \n &=& \lambda_d\vec{J}_d(-t)\cdot \vec{S} +
    \lambda \sin(V(t+i\beta)) \vec{J}_x(-t)\cdot \vec{S} +
    \lambda \cos(V(t+i\beta))\vec{J}_y(-t) \cdot \vec{S}~. \no
\eeqa
The exact form of $\tilde{H}_I$ is actually not important; note
only that it is a linear combination of local operators evolved in
interaction-picture time. Again using the fact that only connected
correlation functions (where $H_I^{(V)}(t)$ and
$\tilde{H}_I^{(V)}(t)$ are connected to $\Or$) occur order by
order in perturbation theory, and the fact that correlation
functions involving $H_I^{(V)}(t)$ and $\tilde{H}_I^{(V)}(t)$
factorize at large times $t$, one can see that all integrals are
convergent in the limit $t_0/\beta\to-\infty$ order by order in
perturbation theory.

Physically, this means that the bath represented by $H_0$ provides
the same mechanism for the steady state to occur as the mechanism
it provides for the system to reach equilibrium in the case $V=0$.

\bigskip

We now cast the expression for the steady state averages in
another suggestive form and derive the alternative formulation,
expressed in (\ref{curssV}), with the steady state obtained by
coupling the dot (i.e. turning on the couplings $\lambda,
\lambda_d$) to leads initially equilibrated at temperature $T$ and
at potential difference $-V$.

Observe that the operators with superscript $(V)$ form the same
current algebra, Eq. (\ref{algxyz}), as those without superscript
since they are obtained by the unitary transformation $U_{V}$
(\ref{trans}):
\beqa
\vec{J}_d^{(V)}(x) &=& U_{-V}\vec{J}_d(x)U_{V}\n
\vec{J}_x^{(V)}(x) &=& U_{-V}\vec{J}_x(x)U_V\n \vec{J}_y^{(V)}(x)
&=& U_{-V}\vec{J}_y(x)U_V\n J_z^{(V)}(x) &=& U_{-V}J_z(x)U_V =
J_z(x) + \frc{V}2
\eeqa
 Hence, the steady-state average of an operator $\Or$
can be written
 \beq
\bra\Or\ket_{ss} = \lim_{t_0/\beta\to-\infty}\frc{\Tr\lt[
\Pexp\lt(i\int_{0}^{t_0}dt\, H_I^{(0)}(t) \rt)\;U_V\, e^{-\beta
H_0} \,U_{-V}\;\Pexp\lt(i\int_{t_0}^0dt\, H_I^{(0)}(t) \rt) \;
U_V\,\Or\, U_{-V} \rt]}{\Tr\lt[e^{-\beta H_0}\rt]}
 \eeq
where we recall that $H_I^{(0)}(t)$ is the operator evolved with
$H_0$ only. Recalling the transformation of $H_0$ under $U_V$
(\ref{transH0}), we find
\beq
\label{opssV} \bra\Or\ket_{ss} = \lim_{t_0/\beta\to-\infty}
\frc{\Tr\lt[ \bar{S}(0,t_0)\; e^{-\beta \lt(H_0-V
H_z\rt)}\;\bar{S}(t_0,0)\; \Or^{(-V)} \rt]}{\Tr\lt[e^{-\beta
\lt(H_0-V H_z\rt)}\rt]}
\eeq
 where
\beq\label
{OrV} \Or^{(-V)} = U_V \Or U_{-V}~.
 \eeq
In (\ref{opssV}), it was necessary to include the factor $e^{\beta
VH_z}$ inside the trace in the denominator of the right-hand side.
For a system on a finite interval, the inclusion of this factor
has the effect of cancelling the constant term that appears in
(\ref{transH0}). Then, the limit of infinite interval is well
defined.

Since for the current operator (\ref{current}) we have
$\cur^{(-V)} = \cur$, this shows the equivalence, for the steady
state current, between the formulation (\ref{opss}) and the
formulation (\ref{curssV}). In general, we will denote the steady
state average in the latter formulation by
\beq\label{opss2}
    \bra\Or\ket_{ss'} \equiv
         \lim_{t_0/\beta\to-\infty} \frc{\Tr\lt[
        \bar{S}(0,t_0)\;
        e^{-\beta \lt(H_0-V H_z\rt)}\;\bar{S}(t_0,0)\;
        \Or \rt]}{\Tr\lt[e^{-\beta \lt(H_0-V
        H_z\rt)}\rt]}~.
\eeq
That is,
\beq
    \bra\Or\ket_{ss'} = \bra U_{-V} \Or U_V \ket_{ss}~.
\eeq

Below, we carry out some formal manipulations which are justified
only if we can establish a more stringent convergence property as
that used above. We need to establish that the following
expression:
\beq
\label{doublelimitssV2} \bra\Or\ket_{ss'} =
\lim_{t_0/\beta\to-\infty} \lim_{t_0'/\beta\to-\infty}
\frc{\Tr\lt[ \bar{S}(0,t_0')\; e^{-\beta \lt(H_0-V
H_z\rt)}\;\bar{S}(t_0,0)\; \Or^{(-V)} \rt]}{\Tr\lt[
\bar{S}(0,t_0')\;e^{-\beta \lt(H_0-V
H_z\rt)}\;\bar{S}(t_0,0)\rt]}~,
\eeq
with the limits on $t_0$ and on $t_0'$ taken independently, will
yield a result independent of the order the limits were taken.
Note that we have included factors $\bar{S}(t_0,0)$ and
$\bar{S}(0,t_0')$ in the denominator. They assure convergence and
cancel by cyclicity of the trace if the limit exists.  To prove
convergence in (\ref{doublelimitssV2}), we consider
\beq
\label{doublelimitssV} \bra\Or\ket_{ss} =
\lim_{t_0/\beta\to-\infty} \lim_{t_0'/\beta\to-\infty}
\frc{\Tr\lt[ \Pexp\lt(i\int_{0}^{t_0'}dt\, H_I^{(V)}(t)\rt)\,
e^{-\beta H_0}\, \Pexp\lt(i\int_{t_0}^0dt\, H_I^{(V)}(t) \rt) \;
\Or \rt]}{\Tr\lt[ \Pexp\lt(i\int_{0}^{t_0'}dt\, H_I^{(V)}(t)\rt)\,
e^{-\beta H_0}\, \Pexp\lt(i\int_{t_0}^0dt\, H_I^{(V)}(t) \rt)\rt]
} ~.
\eeq
Indeed, the same arguments we used to establish the connectedness
and factorization allow us to take, for instance, first the limit
with $|t_0'|$ large, then the limit with $|t_0|$ large, or vice
versa. The result is unique. Using the operator $U_V$ in a manner
similar to the one above, it is a simple matter to obtain
(\ref{doublelimitssV2}) from (\ref{doublelimitssV}) and the result
then is the same as the one obtained from the formulation
(\ref{opss2}).

{\bf Alternative description of the Steady State.} What replaces
the density matrix $e^{-\beta H}$ description of equilibrium? We
could translate the proof establishing equilibrium when ($V =0$)
to the proof establishing steady state when ($V \neq 0$) by means
of the current algebra of symmetries. By similar means we shall
show that a new operator will play for the system in its steady
state a similar role to the one played by the density matrix in
equilibrium. Such a {\em steady-state density matrix} can be
obtained from simple manipulations, now that we have established
the convergence of the integrals.

Consider the formulation (\ref{doublelimitssV2}) of the
steady-state problem, with $\Or$ in (\ref{OrV}) an operator
supported on a finite interval in the theory $H_0$. Bringing
$e^{-\beta H_0}$ completely to the left, the right-hand side of
(\ref{doublelimitssV2}) can be written as follows:
\[
    \lim_{t_0/\beta\to-\infty} \lim_{t_0'/\beta\to-\infty}
        \frc{\Tr\lt[
        e^{-\beta H_0}\,
            \Pexp\lt(i\int_{-i\beta}^{t_0'-i\beta}dt \,
            H_I^{(0)}(t) \rt)\ e^{\beta VH_z}\,
            \Pexp\lt(i\int_{t_0}^{0}dt\,H_I^{(0)}(t) \rt)\
            \Or^{(-V)} \rt]
        }{\Tr\lt[
            e^{-\beta H_0}\,
            \Pexp\lt(i\int_{-i\beta}^{t_0'-i\beta}dt \,
            H_I^{(0)}(t) \rt)\ e^{\beta VH_z}\,
            \Pexp\lt(i\int_{t_0}^{0}dt\,H_I^{(0)}(t) \rt)
         \rt]}~.
\]
Since we showed that the limits can be taken independently, we can
shift $t_0'-i\beta$ to $t_0'$ both in the numerator and in the
denominator, without shifting $t_0$, with vanishing error in the
limit. We can then take $t_0'=t_0$ and keep only one limit symbol.
Inserting
\[
    1 = \Pexp\lt(i\int_{t_0}^{0}dt\,H_I^{(0)}(t) \rt)\,
    \,\Pexp\lt(i\int_{0}^{t_0}dt\,H_I^{(0)}(t) \rt)
\]
just before the operator $e^{\beta VH_z}$ we have,
\[
    \lim_{t_0/\beta\to-\infty}
        \frc{\Tr\lt[
        e^{-\beta H|_{V=0}}\,
            \Pexp\lt(i\int_{0}^{t_0}dt \,
            H_I^{(0)}(t) \rt)\,e^{\beta VH_z}\,
            \Pexp\lt(i\int_{t_0}^{0}dt\,H_I^{(0)}(t) \rt)\
            \Or^{(-V)} \rt]
        }{\Tr\lt[
        e^{-\beta H|_{V=0}}\,
            \Pexp\lt(i\int_{0}^{t_0}dt \,
            H_I^{(0)}(t) \rt)\,e^{\beta VH_z}\,
            \Pexp\lt(i\int_{t_0}^{0}dt\,H_I^{(0)}(t) \rt)
         \rt]}~.
\]
Defining the operator:
\beq\label{Y}
Y =\lim_{t_0/\beta\to-\infty}\bar{S}(0,t_0) H_z \bar{S}(t_0,0)
\eeq
allows us to write the steady-state average of a local operator as
\beq \label{Yave}
\bra\Or\ket_{ss'} = \frc{ \Tr\lt[ e^{-\beta H|_{V=0}}\;e^{\beta
VY} \Or\rt] }{\Tr\lt[ e^{-\beta H|_{V=0}}\;e^{\beta VY} \rt]}~.
\eeq
Note that the limit (\ref{Y}) cannot be expected
to exist as an operator (recall that we are dealing with operators with infinitely many
matrix elements), but only when inserted into appropriate traces (or only when appropriate
matrix elements are considered). More precisely, we have only proven
that (\ref{Y}) is a well-defined operator when it is evaluated in expressions like
(\ref{Yave}), and that the result is the steady-state average of the local
operators inserted. This is a statement solely about a small part of the matrix elements of the operator
(\ref{Y}). The meaning of Eq. (\ref{Yave}) is that one must first evaluate the traces and their ratio
with the expression (\ref{Y}) at finite $t_0/\beta$,
then take the limit indicated in (\ref{Y}) on the result.
The properties of $Y$ as an operator acting in a Hilbert space will be
discussed elsewhere.

Observe, however, that in all situations where the operator $Y$ is well defined (that is, when
we consider the appropriate matrix elements), then it
is a conserved charge. Indeed, when it is well defined, then the
limit $t_0/\beta\to-\infty$ of $S(0,t_0)H_z S(t_0,0)$ (or of any function of this operator) must exist.
Since $S(t_1,t_2) = S(t_1+ dt,t_2+dt)$, we have $\frc{d}{dt}
\lim_{t_0/\beta\to-\infty} S(t,t_0)H_zS(t_0,t)=0$, hence
$[H|_{V=0},Y]=0$. Then we can finally write
\beq\label{opY}
    \bra\Or\ket_{ss'} = \frc{\Tr\lt[e^{-\beta(H|_{V=0}-VY)}
    \Or\rt]}{\Tr\lt[e^{-\beta(H|_{V=0}-VY)}\rt]}~.
\eeq
That is, averages in the steady-state can be obtained by tracing
with an appropriate density matrix\footnote{Observe here that the operator $H|_{V=0}-VY$
need not have a spectrum bounded from below in the Hilbert space of $H|_{V=0}$. As usual, in any explicit
evaluation, the traces are regularized using a regularisation that respects cyclicity.
The resulting expression is well defined with insertion of local operators
only and the limit implied in the definition
of the operator $Y$ should be taken simultaneously in the numerator and the denominator.}.

What difference is there between the equilibrium and the steady
state? Consider a quantum mechanical system described by a
Hamiltonian $H$. Put the system at equilibrium with a bath where
there can be exchange of heat and of any quantity $Q$ that is
conserved by the dynamics $H$. The average of observables is then
described by the density matrix $e^{-\beta(H+\mu Q)}$ where $\mu$
is the chemical potential associated to $Q$: the energy brought to
the system by increasing $Q$ by one unit. In expression
(\ref{opY}), the average of a local operator $\Or$ in the steady
state is a trace with a density matrix of exactly the same form.
The main difference is that the operator $Y$ is a {\em non-local}
conserved charge. A local conserved charge can be written as an
integral over space of a local operator of the theory $H|_{V=0}$
plus a local operator at the impurity site, with possible
non-trivial impurity-space components. The operator $Y$ (\ref{Y})
cannot be written in that way. To see this, we can write it as
follows:
\beq\label{Ycur}
    Y = H_z + \int_{-\infty}^0dt\, \cur(t)
\eeq
where
\beq
    \cur(t) = \bar{S}(0,t)\cur \bar{S}(t,0)
\eeq
is the time-evolved current $\cur$ with respect to the theory
$H|_{V=0}$. Then it is simple to write it as an integral of a
charge density:
\beq
    Y = \int_{-\infty}^\infty dx\, j^{tot}(x,0)
\eeq
with
\beq
    j^{tot}(x,t) = \frc1\pi J_z(x,t) + \delta(x) \int_{-\infty}^t
    dt'\;\cur(t')~.
\eeq
The charge density has a local bulk part, but the term at the
impurity is {\em not} a local field of the theory $H|_{V=0}$: it
is the time integral of the current, and the current is not the
time derivative of a local field.

The non-locality of $Y$ is the main difference between the
description of a steady state and of an equilibrium state. In the
formulation (\ref{opY}), only a restricted set of operators $\Or$
have well-defined average: those that have {\em stationary}
expectation values. All local operators are of this type, but, for
instance, it is simple to see that the operator $H_z$ does not
have a well-defined steady-state value.

Note also that the operator $Y$ gives in principle a description
of the {\em asymptotic state} that one can use in order to
describe the steady state: quantities in the steady state can be
evaluated as averages in an appropriate asymptotic state. Further
analysis in this direction will be presented in our future works.

We wish to remark that sometime ago  Hershfield
\cite{Hershfield93} has considered steady-state flow and has
argued that under some assumptions concerning the relaxation of
correlation function an expression (\ref{opY}) would govern the
steady-state current. He gave then implicit equations to determine
$Y$. It appears to us that our explicit expressions for the
operator $Y$ satisfies his implicit equations, and should hence
correspond to the same operator (although we have not thoroughly
ascertained the confluence of the two approaches). We must stress,
however, that no assumptions were made in our derivation.

\section{RG-improved real-time perturbation theory}

{\bf The perturbative expansion.} We now turn to real-time
perturbation theory for the current (\ref{curssV}). We take the
formulation where the system is initially brought to equilibrium
with a nonzero bias voltage, then disconnected from the external
bath before the voltage is turned off and the interaction is
turned on. It will be convenient to consider adiabatically turning
on the interaction in the infinite past: we introduce a large-time
exponential cutoff, $e^{\eta t} H_I$, with  $\eta$ a positive
scale with dimension of energy, and take the limit
$t_0/\beta\to-\infty$ in (\ref{curssV}). The quantity $\eta\beta$
will be sent to 0 at the end of the calculations. This means,
physically, that the two leads are slowly brought towards the dot
after the voltage has been turned off. Our proof that there are no
divergencies as $t_0/\beta\to-\infty$ in the previous section
shows that there are no divergencies as $\eta\beta \to 0$. The
current can then be written
\beq
    \bra\cur\ket_{ss} =
    \lim_{\eta\beta\to0^+}
    \frc{\Tr\lt[e^{-\beta(H_0-VH_z)}\,S_\eta(-\infty,0)\cur
    S_\eta(0,-\infty)\rt]}{\Tr\lt[e^{-\beta(H_0-VH_z)}\rt]}
\eeq
with \footnote{By definition, $dS_\eta(t_1,t_2)/dt_2 =
iS_\eta(t_1,t_2)(H_0 + e^{\eta t_2} H_I)$ and $S_\eta(t,t)=1$.}
\beq
    S_\eta(t_1,t_2) = \Pexp\int_{t_1}^{t_2} i \lt(H_0 + e^{\eta t} H_I\rt)
    \, dt~.
\eeq
More precisely,
\beq\label{rtpert}
    \bra\cur\ket_{ss} = \lim_{\eta\beta\to0^+}
    \sum_{k=0}^\infty i^k \int_{-\infty}^0
    \hspace{-5mm}dt_1 e^{\eta t_1}\hspace{-2mm}\int_{t_1}^0 \hspace{-2mm}dt_2 e^{\eta t_2} \cdots
    \int_{t_{k-1}}^0 \hspace{-5mm}dt_k e^{\eta t_k}
    \bra\bra[H_I(t_1),[H_I(t_2),\cdots,[H_I(t_k),\cur]\cdots]]\ket\ket_V
\eeq
where
\beq
    \bra\bra \cdots \ket\ket_V = \frc{\Tr\lt[e^{-\beta \lt(H_0-V H_z\rt)}
    \cdots\rt]}{\Tr\lt[e^{-\beta \lt(H_0-V H_z\rt)}\rt]}~.
\eeq

The integrals in this expansion are plagued with ultraviolet
divergencies which we have to regularize. This can be done in
several ways. For our purposes, it will be convenient to modify
the operators $\vec{J}_d(x),\,\vec{J}_x(x),\,\vec{J}_y(x)$ in
order to render their correlation functions regular at coinciding
points. More precisely, we choose the regularization scheme where
all operators (in the Hamiltonian and in correlation functions) at
the impurity site are regularized, whereas all operators away from
it are unaffected. Since the interaction is only at the impurity
site, this is enough to regularize the theory. Define the momentum
space (mode) operators
$\vec{J}_d(p),\,\vec{J}_x(p),\,\vec{J}_y(p)$ and $J_z(p)$ in the
following way:
\beqa
&&    \vec{J}_d(x) = \int_{-\infty}^{\infty} dp \,
    \vec{J}_d(p)e^{ipx}~,\quad
    \vec{J}_x(x) = \int_{-\infty}^{\infty} dp \,
    \vec{J}_x(p)e^{ipx}~, \no\\
&&   \vec{J}_y(x) = \int_{-\infty}^{\infty} dp \,
    \vec{J}_y(p)e^{ipx}~,\quad
    J_z(x) = \int_{-\infty}^{\infty} dp \,
    J_z(p)e^{ipx}~.\no
\eeqa
The mode operators satisfy the following set of commutation
relations:
\beqa\label{algxyzmodes}
    [J_d^i(p),J_d^j(q)] &=& i \;\epsilon_{ijk}\; J_d^k(p+q)
     + p \; \delta_{ij} \;\delta(p+q)
    \no\\ {}
    [J_x^i(p),J_x^j(q)] &=& i \; \epsilon_{ijk} \; J^k_d(p+q)
     + p \;\delta_{ij} \;\delta(p+q)
    \no\\ {}
    [J_y^i(p),J_y^j(q)] &=& i \; \epsilon_{ijk} \; J^k_d(p+q)
     + p \;\delta_{ij} \;\delta(p+q)
    \no\\ {}
    [J_z(p),J_z(q)] &=&  p \;\delta(p+q)
    \no\\ {}
    [J_d^i(p),J_x^j(q)] &=& i \;\epsilon_{ijk} \; J^k_x(p+q)
    \\ {}
    [J_d^i(p),J_y^j(q)] &=& i \;\epsilon_{ijk} \; J^k_y(p+q)
    \no\\ {}
    [J_d^i(p),J_z(q)] &=& 0
    \no\\ {}
    [J_x^i(p),J_y^j(q)] &=& i \;\delta_{ij} \; J_z(p+q)
    \no\\ {}
    [J_y^i(p),J_z(q)] &=& i \;J_x^i(p+q)
    \no\\ {}
    [J_z(p),J_x^i(q)] &=& i \;J_y^i(p+q)~.
    \no {}
\eeqa
We then introduce the regularized operators
\beqa
&&    (\vec{J}_d)_\Lambda(x) = \int_{-\infty}^{\infty} dp \,
    R_\Lambda(p)\,\vec{J}_d(p)e^{ipx}~,\quad
    (\vec{J}_x)_\Lambda(x) = \int_{-\infty}^{\infty} dp \,
    R_\Lambda(p)\,\vec{J}_x(p)e^{ipx}~, \no\\
&&    (\vec{J}_y)_\Lambda(x) = \int_{-\infty}^{\infty} dp \,
    R_\Lambda(p)\,\vec{J}_y(p)e^{ipx}\no
\eeqa
where $R_\Lambda(p)$ is a function that vanishes as
$|p|\to\infty$, and whose complex conjugate satisfies $
R_\Lambda(p)^* = R_\Lambda(-p) $ in order to preserve hermiticity
of the regularized operators. The function $R_\Lambda(p)$ can be
chosen in many ways, and the choice is a matter of convenience. We
will choose a gaussian regularization,
\beq\label{regLambda}
    R_\Lambda(p) = e^{-p^2/(2\Lambda^2)}~.
\eeq
The parameter $\Lambda$ plays the role of an effective band width
(we do not denote it $D$ since it is certainly not exactly the
band width). The universal part of the limit $\Lambda\gg V,T$ is
the same as that of the limit of large band width.

The time integrals in (\ref{rtpert}) can now be traded to momentum
integrals:
\beqa
    \bra\cur\ket_{ss} &=& \sum_{k=0}^\infty (-1)^k \int
    \frc{dp_1\,R_\Lambda(p_1)}{p_1+i\eta} \int \frc{dp_2\,R_\Lambda(p_2)}{p_1+p_2+2i\eta}  \cdots
    \int \frc{dp_k\,R_\Lambda(p_k)}{p_1+p_2+\cdots+p_k+ki\eta} \no\\
&& \qquad\qquad\qquad\int
    ds\,R_\Lambda(s)\;\bra\bra[\tilde{H}_I(p_1),[\tilde{H}_I(p_2),\cdots,
        [\tilde{H}_I(p_k),\tilde{\cur}(s)]\cdots]]\ket\ket_V
\eeqa
where
\beq
    \tilde{H}_I(p) = \lambda_d\vec{J}_d(p)\cdot \vec{S} +
    2 \vec{J}_y(p)\cdot \vec{S}
\eeq
and
\beq
    \tilde{\cur}(p) = \vec{J}_x(p)\cdot S~.
\eeq
The parameter $\eta$ can be set to zero with the requirement that
the momentum integrals be taken on a line parallel to the real
axis in the $p$-plane  with a slight positive imaginary part:
\beqa
    \bra\cur\ket_{ss} &=& \sum_{k=0}^\infty (-1)^k \int_+
    \frc{dp_1\,R_\Lambda(p_1)}{p_1} \int_+ \frc{dp_2\,R_\Lambda(p_2)}{p_1+p_2}  \cdots
    \int_+ \frc{dp_k\,R_\Lambda(p_k)}{p_1+p_2+\cdots+p_k} \no\\
&& \qquad\qquad\quad\int
    dq\,R_\Lambda(q)\;\bra\bra[\tilde{H}_I(p_1),[\tilde{H}_I(p_2),\cdots,
        [\tilde{H}_I(p_k),\tilde{\cur}(q)]\cdots]]\ket\ket_V~.
    \label{rtpertmom}
\eeqa
Note that this can be done since $R_\Lambda(p)$ does not have a
singularity at $p=0$ or anywhere on the real $p$-line. No
singularity at $p_1+\cdots+p_j=0$ (for $j=1,\ldots,k$) can occur
in the averages $\bra\bra \cdots\ket\ket_V$ in the expression
above, since we know that there are no divergencies as
$\eta\beta\to0$.

In fact, for explicit calculations, it will be more convenient to
integrate over the real line in momentum space, and to use the
formalism of principal value integrals. For this purpose, recall
that
\beq\label{prinval}
    \int_+ d p \frc{f(p)}{(p+q)^n} = \int d p \lt(\frc{(-1)^{n}}{(n-1)!}i\pi\delta^{(n-1)}(p+q)
    + \prin \frc1{(p+q)^n}\rt) f(p)
\eeq
where $\prin$ says that we must take the principal value of the
integral:
\beq
    \int d p \lt(\prin \frc{1}{(p+q)^n}\rt) f(p) =
    \mbox{finite part of}\ \lt(\int_{-\infty}^{-\epsilon} + \int_{\ep}^\infty\rt)
    d p \frc{f(p)}{(p+q)^n} \ \mbox{in power expansion as $\ep\to0$}~.
\eeq

{\bf Explicit one-loop calculations.} The trace over the impurity space of the operators
\[
    [\tilde{H}_I(p_1),[\tilde{H}_I(p_2),\cdots,
        [\tilde{H}_I(p_k),\tilde{\cur}(q)]\cdots]]
\]
can be obtained by using the following general formula, valid for
any vector operators $\vec{A}$ and $\vec{B}$ that commute with
$\vec{S}$:
\beq
    [\vec{A}\cdot \vec{S},\vec{B}\cdot \vec{S}] =
    \frc{i}2\,\{A^i,B^j\}\,\epsilon_{ijk}\,S^k + \frc14 [A^i,B^i]
\eeq
where $\{\cdot,\cdot\}$ is the anti-commutator. Using the
commutation relations (\ref{algxyzmodes}) for evaluating the
commutators, one is left then only with multiple anti-commutators
of the mode operators. To one loop we will need the following
commutators:
\beqa
    C_1 &=& [\vec{J}_y(q)\cdot\vec{S},\vec{J}_x(k)\cdot\vec{S}] \n
    C_2 &=&
    [\vec{J}_d(p)\cdot\vec{S},[\vec{J}_y(q)\cdot\vec{S},\vec{J}_x(k)\cdot\vec{S}]]
    \n
    C_3 &=&
    [\vec{J}_y(p)\cdot\vec{S},[\vec{J}_d(q)\cdot\vec{S},\vec{J}_x(k)\cdot\vec{S}]]~.
\eeqa
Under trace over the impurity space, they give
\beqa
    \Tr_{{\rm impurity}}(C_1) &=& -3i J_z(q+k) \n
    \Tr_{{\rm impurity}}(C_2) &=& 2\{J_y^i(q), J^i_x(p+k)\} - 2\{J_y^i(p+q),J^i_x(k)\}
    \n
    \Tr_{{\rm impurity}}(C_3) &=& - 2\{J_y^i(p+q),J^i_x(k)\}
    \label{TrC}~.
\eeqa
The traces $\bra\bra \cdots \ket\ket_V$ over the bulk-CFT Hilbert
space of these anti-commutators can be calculated without the need
to construct the Hilbert space for $H_0-V H_z$, but only by using
the following exchange relations:
\beqa
  \vec{J}_d(p) \,e^{-\beta\lt(H_0-V H_z\rt)} &=&
    e^{-\beta\lt(H_0-V H_z\rt)} \,
    \vec{J}_d(p)\,e^{-\beta p}\no\\
    J_z(p) \,e^{-\beta\lt(H_0-V H_z\rt)} &=&
    e^{-\beta\lt(H_0-V H_z\rt)} \,
    J_z(p)\,e^{-\beta p}\no\\
  \vec{J}_+(p) \,e^{-\beta\lt(H_0-V H_z\rt)} &=&
    e^{-\beta\lt(H_0-V H_z\rt)} \,
    \vec{J}_+(p)\,e^{-\beta (p+V)} \no\\
    \vec{J}_-(p) \,e^{-\beta\lt(H_0-V H_z\rt)} &=&
    e^{-\beta\lt(H_0-V H_z\rt)} \,
    \vec{J}_-(p)\,e^{-\beta (p-V)}
\eeqa
where we use the following linear combinations:
\beq
    \vec{J}_+(p) = \frc12\lt(\vec{J}_x(p)+i\vec{J}_y(p)\rt)~,\quad
    \vec{J}_-(p) = \frc12\lt(\vec{J}_x(p)-i\vec{J}_y(p)\rt)~.
\eeq
The exchange relations imply for any operator $\Or$
\beqa
  \bra\bra \Or \vec{J}_d(p) \ket\ket_V &=&
    e^{-\beta p}\,\bra\bra \vec{J}_d(p) \Or \ket\ket_V \no\\
    \bra\bra \Or J_z(p) \ket\ket_V &=&
    e^{-\beta p}\, \bra\bra J_z(p) \Or \ket\ket_V\no\\
  \bra\bra \Or \vec{J}_+(p) \ket\ket_V &=&
    e^{-\beta (p+V)}\,\bra\bra \vec{J}_+(p) \Or \ket\ket_V
    \no\\
    \bra\bra \Or \vec{J}_-(p) \ket\ket_V &=&
    e^{-\beta (p-V)}\,\bra\bra \vec{J}_-(p) \Or \ket\ket_V~.
\eeqa
Using the commutator formulas (\ref{algxyzmodes}), using these
relations and using the traces of single operators, all traces can
be calculated. Traces of single operators can be calculated from
the fact that $\bra\bra \vec{J}_{d,x,y}\ket\ket_0 = \bra\bra J_z
\ket\ket_0=0$, and from $\bra\bra \Or\ket\ket_V = \bra\bra
U_{-V}\Or U_V \ket\ket_0$. In terms of modes, the operator
$U_{-V}$ has the representation
\beq\label{UVmode}
    U_{-V} = e^{2VJ_z'(0)}
\eeq
where we formally define $J_z'(p) = dJ_z(p)/dp$. The traces of
single operators are then given by
\beq
    \bra\bra\vec{J}_{d,x,y}(p)\ket\ket_V = 0~,\quad \bra\bra
    J_z(p)\ket\ket_V = V \delta(p)~.
\eeq
The trace of the only type of anti-commutator appearing in
(\ref{TrC}) is then given by
\beq \label{traces}
    \bra\bra \{J_y^i(p),J_x^j(q)\}\ket\ket_V = -\frc{i}8 F(p)
    \delta(p+q)\delta_{ij}
\eeq
where
\beq
    F(p) = (p+V)\,\frc{1+e^{-\beta (p+V)}}{1-e^{-\beta (p+V)}} -
            (p-V)\,\frc{1+e^{-\beta (p-V)}}{1-e^{-\beta (p-V)}}~.
\eeq
From (\ref{rtpertmom}) and using (\ref{prinval}), the integrals to
be calculated at one loop are
\beqa
    I_1 &=&  -\int
    dq\,R_\Lambda(q)\lt(-i\pi\delta(q) + \prin\frc1{q}\rt)
    \int dk\,R_\Lambda(k)\;\bra\bra C_1\ket\ket_V \n
    I_{2,3} &=& \int
    dp\,R_\Lambda(p)\lt(-i\pi\delta(p) + \prin\frc1{p}\rt)
    \int dq\,R_\Lambda(q)\lt(-i\pi\delta(p+q) + \prin\frc1{p+q}\rt)
\times\no\\
&& \qquad\qquad\quad\int
    dk\,R_\Lambda(k)\;\bra\bra C_{2,3}\ket\ket_V~.
    \label{integralsC}
\eeqa
The current will then be given by
\beq\label{curI}
    \bra\cur\ket_{ss} = \la^2\lt(
    I_1 + \ld (I_2+I_3) +  O(\ld^2,\la^2)\rt) ~.
\eeq
Since all integrals are real, it is clear that in the expressions
(\ref{integralsC}), only terms with an odd number of delta
functions in the momentum variables give non-zero contributions.
It is a simple matter then to obtain
\beqa\label{I1}
    I_1 &=& \frc{3\pi V}2 \\
    I_2 = I_3 &=& \frc{3\pi V}4 \int d p \, \prin\frc1p f(p)\,R_\Lambda(Vp)^2
\eeqa
where $p$ is now a dimensionless momentum variable, and
\beq\label{fp}
    f(p) = (p+1)\,\frc{1+e^{-w(p+1)}}{1-e^{-w(p+1)}} -
            (p-1)\,\frc{1+e^{-w(p-1)}}{1-e^{-w(p-1)}}
\eeq
and
\beq
    w = \beta V~.
\eeq
Using the symmetry $f(-p) = -f(p)$, the integral $I_2$ can be
calculated in the following way, keeping only the divergent and
finite parts as $\Lambda\to\infty$:
\beqa
    I_2 = I_3 &=& \frc{3\pi V}2 \int_0^\infty \frc{d p}p\,
    f(p)\,e^{-p^2V^2/\Lambda^2} \n
        &\sim&
        \frc{3\pi V}2 \lt[ \int_0^\infty \frc{d p}p \,
        (f(p)-2+2e^{-p^2}) -
        2\int_0^\infty \frc{d p}p \,(e^{-p^2}-1)
        \,e^{-p^2V^2/\Lambda^2}\rt] \n
        &\sim&
        3\pi V \lt[ P(w) + \ln\lt(\frc{\Lambda}V\rt)\rt]
        \label{I2}
\eeqa
where the symbol $\sim$ means that equality is valid only for the
finite and divergent parts as $\Lambda\to\infty$ and where
\beq\label{Pw}
    P(w) = \frc12 \int_0^\infty \frc{d p}p \,
        (f(p)-2+2e^{-p^2}) = \int_{0}^{\infty} \frc{d p}p\lt(\frc{p+1}{e^{w(p+1)}-1} -
    \frc{p-1}{e^{w(p-1)}-1} + e^{-p^2}\rt)~.
\eeq
This is a well-defined function for all $w$ with positive real
part, and is easy to evaluate numerically.

It will be useful to have the asymptotic behavior of the function
$P(w)$, at large and small $w$. This is evaluated in Appendix \ref{appasympP}.
The asymptotic expansion at large $w$ is given by
\beq\label{Plargew}
    P(w) - 1 + \frc{\gamma}2 \sim -\frc{\pi^2}3 w^{-2} \quad
    \mbox{as} \quad w\to\infty~,
\eeq
whereas the expansion at small $w$ is
\beq\label{Psmallw}
    P(w) = \ln(w) + \frc{\sqrt{\pi}}2 +\gamma-\ln(2\pi) + O(w) \quad\mbox{as}\quad w\to0~.
\eeq


{\bf Summarizing}, the zero- and one-loop divergent and finite
contributions to the current are (with $T=1/\beta$)
\beq\label{curoneloop}
    \bra\cur\ket_{ss} = 3\pi V\, \lambda^2\lt[\frc12+2
    \lambda_d\lt(P\lt(\frc{V}T\rt)+\ln\lt(\frc{\Lambda}{V}\rt)\rt) \rt]
\eeq
with $P(w)$ given by the integrals in (\ref{Pw}), $P(w) =
\int_{0}^{\infty} \frc{d p}p\lt(\frc{p+1}{e^{w(p+1)}-1} -
    \frc{p-1}{e^{w(p-1)}-1} + e^{-p^2}\rt)$. Note
that only the combination $P(V/T)+\ln(\Lambda/V)$ appears at one
loop. This combination has the limits
\beqa
    P\lt(\frc{V}T\rt)+\ln\lt(\frc{\Lambda}V\rt) &\sim&
        - \lt(\ln\lt(\frc{V}{\Lambda}\rt)  - 1 + \frc{\gamma}2 \rt)
\quad \quad  \quad   \quad  \quad \quad\mbox{as}\quad T \ll V \ll
        \Lambda \n
    &\sim& - \lt(\ln\lt(\frc{T}{\Lambda}\rt) - \frc{\sqrt{\pi}}2 - \gamma+\ln(2\pi) \rt)
    \quad\mbox{as}\quad V\ll T\ll \Lambda~.
\eeqa
The one-loop calculation was also performed in \cite{KaminskiNG00} (although the analysis
did not go in as much detail as ours), and it can be verified that their results agree with ours.

{\bf Results for the two-loop calculations.} Two-loop integrals
come from the commutators (here $p$, $q$, $r$ and $k$ are all
momentum variables):
\beqa
    C_4 &=&
    [\vec{J}_d(p)\cdot\vec{S},[\vec{J}_d(q)\cdot\vec{S},
    [\vec{J}_y(r)\cdot\vec{S},\vec{J}_x(k)\cdot\vec{S}]]] \n
    C_5 &=&
    [\vec{J}_d(p)\cdot\vec{S},[\vec{J}_y(q)\cdot\vec{S},
    [\vec{J}_d(r)\cdot\vec{S},\vec{J}_x(k)\cdot\vec{S}]]] \n
    C_6 &=&
    [\vec{J}_y(p)\cdot\vec{S},[\vec{J}_d(q)\cdot\vec{S},
    [\vec{J}_d(r)\cdot\vec{S},\vec{J}_x(k)\cdot\vec{S}]]] \n
    C_7 &=&
    [\vec{J}_y(p)\cdot\vec{S},[\vec{J}_y(q)\cdot\vec{S},
    [\vec{J}_y(r)\cdot\vec{S},\vec{J}_x(k)\cdot\vec{S}]]]
\eeqa
appearing inside the traces in the integrands of
(\ref{rtpertmom}). Calculating the traces and using
(\ref{prinval}), we can find the corresponding set of two-loop
integrals, $I_4, I_5, I_6$ and $I_7$ (written in Appendix \ref{appintegtwoloop}).
These integrals enter the current as
\beq\label{curI2}
    \bra\cur\ket_{ss} = \la^2\lt(
    I_1 + \ld (I_2+I_3) +  \ld^2 (I_4+I_5+I_6) + \la^2 I_7 + O(\ld^3,\la^2\ld)\rt) ~.
\eeq
Explicit calculation of these integrals lead to the following
divergent parts as $\Lambda\to\infty$:
\beqa
    I_4+I_5+I_6 &\sim& 3\pi V
    \lt[10P(w)+5\ln\lt(\frc{\Lambda}V\rt)-1\rt]\ln\lt(\frc{\Lambda}V\rt)
    + {\rm finite}
    \n
    I_7 &\sim& 3\pi V \lt[2P(w)+\ln\lt(\frc{\Lambda}V\rt)-1\rt]\ln\lt(\frc{\Lambda}V\rt)
    + {\rm finite}~.
    \label{2loopdiv}
\eeqa
The finite contributions are much more complicated, and are
reported in Appendix \ref{appfinitetwoloop}. Let us only notice that the limit
$T/V\to0$ of these contributions is finite, as in the one-loop
results. This indicates that the voltage $V$ plays the role of a
good infrared cutoff for the perturbative calculation of the
current to two loops: the temperature may be set to zero without
divergences to this order.

{\bf Summarizing}, the zero-, one- and two-loop finite and
divergent contributions to the current are,
\beqa
    \bra\cur\ket_{ss} &=& 3\pi V\, \lambda^2\lt\{\frc12 + 2\lambda_d\lt(P(w) +
    \ln\lt(\frc{\Lambda}V\rt)\rt) + \rt.\n && \qquad\quad\lt.
    + \lambda_d^2
    \lt[10P(w)+5\ln\lt(\frc{\Lambda}V\rt)-1\rt]\ln\lt(\frc{\Lambda}V\rt)
    +\lambda^2
    \lt[2P(w)+\ln\lt(\frc{\Lambda}V\rt)-1\rt]\ln\lt(\frc{\Lambda}V\rt)
    \rt\} + \n && + \lambda^2\lambda_d^2 \,
    [I_4+I_5+I_6]_{\rm finite} + \lambda^4\,[I_7]_{\rm finite}
\eeqa
where $[I_4+I_5+I_6]_{\rm finite}$ and $[I_7]_{\rm finite}$ are
given in Appendix \ref{appfinitetwoloop}.

{\bf The Renormalization Group equation (The Callan-Symanzik
equation).} In systems at equilibrium, Wilson's renormalization
group ideas allow us to understand how physical quantities can
have universal forms (independent of the precise form of the
interactions at the microscopic level) when all physical energy
scales (temperature, voltage, etc.) are much lower than the
microscopic scales (band width, inverse lattice spacing, etc.).
Out of equilibrium, it is not obvious that Wilson's
renormalization group ideas still apply.

A different, but equivalent, way to look at universality is to
study the limit where the cut-off $\Lambda$ is increased and sent
to infinity, while the couplings dependence on the cut-off is
again governed by an RG equation, valid at very large cut-off. If
such a limit exists then all quantities tend to their universal
form.

We will argue that the steady-state average of the current
operator (or of any operator having a well-defined average in the
steady state) satisfies the Callan-Symanzik equation with the same
beta function and anomalous dimension as they occur in any average
evaluated at equilibrium. More precisely, we will argue that
\beq\label{CS}
    \lt(\Lambda\lt.\frc{\p}{\p \Lambda}\rt|_{\lambda,\lambda_d}
    +\beta_\lambda(\lambda,\lambda_d) \frc{\p}{\p\lambda}
    +\beta_{\lambda_d}(\lambda,\lambda_d) \frc{\p}{\p\lambda_d}\rt)
    \bra\cur\ket_{ss} \stackrel{\Lambda\to\infty}=
    0
\eeq
where $\beta_\lambda$ and $\beta_{\lambda_d}$ are the beta
functions of the anisotropic two-channel Kondo model. Note that
the anomalous dimension term does not occur: the current operator
$\cur$ has zero anomalous dimension (this is natural from a
physical perspective, as the current is a physical object which
should not change with a change of scales; we will verify this
explicitly to one loop, and indirectly to two loops, below).

We should note that the Callan-Symanzik equations with one-loop beta functions and zero anomalous dimension
was written in \cite{KaminskiNG00} for the steady-state average of the current, from
physical arguments\footnote{The authors also considered the case of an oscillating applied voltage,
where they had to introduce an extra scale, a ``decoherence time,''
in order to recover universality. We do not consider this case here.}.
Here we present a quantum field theoretic argument that applies
to all orders (and all matrix elements), and in the next section we explicitly verify this argument and
calculate the beta functions and anomalous dimension in a universal fashion (so that
it automatically applies to the steady state) to one-loop order.

The Callan-Symanzik equation embodies Wilson's renormalization
group ideas: it tells us how a change of cutoff $\Lambda$ (for
instance, the band width) can be compensated by a change of few
relevant coupling constant, as long as all physical energy scales
are much lower than $\Lambda$. Solving the Callan-Symanzik
equations (this is done below for the current in the steady state)
allows us to describe the low energy behavior of the steady-state
current in terms of the ratios $V/T_K$ and $T/T_K$, where $T_K$ is
an integration constant, as well as of one extra parameter
(invariant under the RG flow) characterizing the asymmetry between
the couplings $\lambda_d$ and $\lambda$; we will denote this
parameter by $C$. The integration constant $T_K$ and the extra
parameter $C$ characterize the quantum field theory; when they are
fixed, all averages can be evaluated unambiguously. These
parameters are not universal: different microscopic theories have
low-energy behaviors described by different values for them. Up to
these non-universal quantities, the quantum field theory
description is universal, independent of the precise choice of the
cutoff procedure (precise structure of the band, for instance) and
of irrelevant couplings (interactions that give vanishing
contributions at low energies). The integration constant $T_K$ is
the Kondo temperature: the temperature above which the ``Kondo
cloud'' gets destroyed by the thermal energy. It is related to the
microscopic values of the couplings $\lambda$ and $\lambda_d$ (the
values when $\Lambda$ is chosen to be of the order of the real
band width), and it decreases if the couplings are decreased. At
zero couplings, the Kondo temperature is zero and all scales of
the low-energy physics disappear: this is a quantum critical
point. The quantum field theory with finite ratios $V/T_K$ and
$T/T_K$ describes the situation $T_K\ll\Lambda$: the couplings are
sent to zero at the same time as the voltage and temperature are
made much smaller than $\Lambda$. This is the scaling limit,
describing the region around the quantum critical point
$\lambda=\lambda_d=0$. In particular, one finds that the
perturbative expansion is valid in the region $T_K \ll V$ or
$T_K\ll T$.

We will see that in the steady state, it will be more convenient
to introduce a scale $M_K$ characterizing both the effects of the
thermal energy and of the electric potential driving the current
on the Kondo cloud. We will then be able to compare these effects,
using a comparison parameter that is exact in one-loop
perturbation theory.

It is natural that the Callan-Symanzik equation is still valid in
the steady state, since the steady state can be understood as an
appropriate asymptotic state, characterized by a scale $V$, and
since averages of operators in asymptotic states satisfy the
Callan-Symanzik equation. In particular, $V$ should flow trivially
with the renormalization group. However, it is instructive to see
explicitly how this works in real-time perturbation theory.

We will argue that Eq. (\ref{CS}) holds simply from the fact that
the Callan-Symanzik equation is satisfied for any average in
equilibrium. The main observation is the following. Consider the
Hilbert space ${\cal H}_{-V}$ associated to the Hamiltonian
$H_0-VH_z$. It is not formed of vectors that are in the Hilbert
space ${\cal H}_{0}$ associated to the Hamiltonian $H_0$. In
particular, its ground state $|-V\ket$ can be formally defined as
\[
    U_V|0\ket~,
\]
where $|0\ket$ is the ground state of $H_0$. This definition
indeed makes sense for any finite length of the system, and in
fact allows to calculate matrix elements of any operators also at
infinite length, but it does not give a vector in ${\cal H}_0$ at
infinite length. Nevertheless, the mode operators associated to
current algebra operators still have a well-defined action on
${\cal H}_{-V}$. In order to see this, it is convenient to
construct the Hilbert space ${\cal H}_0$ by defining a vacuum
state $|0\ket$ satisfying $\vec{J}_{+,-,d}(p)|0\ket=0$ and
$J_z(p)|0\ket=0$ for all $p\ge0$, and by constructing other states
of the Hilbert space by acting with $\vec{J}_{+,-,d}(p)$ and
$J_z(p)$ at $p<0$. Then, it is a simple matter to see, using
(\ref{UVmode}), that
\beqa
    \vec{J}_+(p) |-V\ket &=& 0 \quad \mbox{if and only if}\quad  p\ge-V \n
    \vec{J}_-(p) |-V\ket &=& 0 \quad \mbox{if and only if}\quad  p\ge V \n
    \vec{J}_d(p) |-V\ket &=& 0 \quad \mbox{if and only if}\quad  p\ge0 \n
    J_z(p) |-V\ket &=& 0 \quad \mbox{if and only if}\quad  p>0~.
\eeqa
The main observation is that any normal ordering operation valid
on ${\cal H}_0$ {\em is still} a good normal ordering operation on
${\cal H}_{-V}$. Indeed, a state of the form
\[
    J_+^{i_1}(p_1) \cdots J_+^{i_a}(p_a)\
    J_-^{j_1}(p_1') \cdots J_-^{j_b}(p_b')\
    J_d^{k_1}(p_1'') \cdots J_d^{k_c}(p_c'')\
    J_z(p_1''') \cdots J_z(p_d''')\ |-V\ket
\]
gives zero whenever
\[
    p_1+\ldots+p_a+p_1'+\ldots+p_b'+p_1''+\ldots+p_c''+p_1'''+\ldots+p_d'''+
    (a-b)V > 0~.
\]
Since $(a-b)V$ is a much smaller than $\Lambda$ for any finite $a$
and $b$, a normal ordering of the type
\beq\label{normord}
    : J(p_1) J(p_2) \cdots J(p_k) : = J(p_{i_1})J(p_{i_2})\cdots J(p_{i_k})
    \quad \mbox{with}\quad p_{i_1}\le p_{i_2}\le\cdots\le
    p_{i_k}~,
\eeq
where the indices $i_m$'s are all different and drawn in
appropriate fashion from the set $\{1,2,\ldots,k\}$ as to make the
set of inequalities for the $p_{i_m}$'s valid, is still a good
normal ordering on ${\cal H}_{-V}$.

This observation is enough in order to see that the
Callan-Symanzik equation is still valid in the steady state.
Indeed, the Callan-Symanzik equation is really an equation for
operators, rather than just for particular averages. Recalling the
regularized real-time perturbation theory (\ref{rtpertmom}),
consider the operator
\beqa
&&    \bar\cur_\Lambda = \sum_{k=0}^\infty (-1)^k \int
    dp_1\,R_\Lambda(p_1)\lt(-i\pi\delta(p_1) + \prin\frc1{p_1}\rt)
    \int dp_2\,R_\Lambda(p_2)\lt(-i\pi\delta(p_1+p_2) + \prin\frc1{p_1+p_2}\rt) \times\no\\
&&  \qquad\qquad\quad  \cdots \times
    \int dp_k\,R_\Lambda(p_k)
        \lt(-i\pi\delta(p_1+p_2+\cdots+p_k) + \prin\frc1{p_1+p_2+\cdots+p_k}\rt)
    \,\times\no\\
&& \qquad\qquad\quad\int
    ds\,R_\Lambda(s)\;[\tilde{H}_I(p_1),[\tilde{H}_I(p_2),\cdots,
        [\tilde{H}_I(p_k),\tilde{\cur}(s)]\cdots]]~.
    \label{barcur}
\eeqa
It gives the interacting current operator with respect to the free
Hilbert space ${\cal H}_0$ (at equilibrium) or ${\cal H}_{-V}$ (in
the steady state). The validity of the Callan-Symanzik equations
at equilibrium means that all matrix elements of
$\bar\cur_\Lambda$ in the Hilbert space ${\cal H}_0$ satisfy the
Callan-Symanzik equations. This can be written:
\beq\label{CSoper}
    \lt(\Lambda\lt.\frc{\p}{\p \Lambda}\rt|_{\lambda,\lambda_d}
    +\beta_\lambda(\lambda,\lambda_d) \frc{\p}{\p\lambda}
    +\beta_{\lambda_d}(\lambda,\lambda_d) \frc{\p}{\p\lambda_d}\rt)
    \bar\cur_\Lambda \stackrel{\Lambda\to\infty}=
    0 \quad \mbox{on} \quad {\cal H}_0~.
\eeq
Indeed, such matrix elements correspond, in perturbation theory,
to matrix elements of the current operator $\cur$ in the basis of
eigenstates of the full Hamiltonian $H|_{V=0}$ (all of which obey
the same Callan-Symanzik equation). We give an argument for the
validity of (\ref{CSoper}) in Appendix \ref{appcalsym}. In (\ref{CSoper}), the
limit $\Lambda\to\infty$ should be performed after a matrix
element has been calculated, holding the momenta associated to
this matrix element
 fixed. We now extract the divergent and finite part of
the operator $\bar\cur$ as $\Lambda\to\infty$:
\beq\label{curexp}
    \bar\cur_\Lambda \sim \sum_{j=0}^\infty \frc1{j!}\ln^j(\Lambda) :\Or_j:
\eeq
where $\Or_j$'s are operators built out of the mode operators for
the current algebra along with possible impurity-space operators
(the normal ordering is defined in (\ref{normord})). This is
always possible to do by rewriting $\bar\cur_\Lambda$ in terms of
normal-ordered operators and evaluating the coefficients as
$\Lambda\to\infty$. The Callan-Symanzik equation (\ref{CSoper})
then
 states,
\beq\label{CSrecurs}
    :\Or_{j+1}: +   \lt(\beta_\lambda(\lambda,\lambda_d) \frc{\p}{\p\lambda}
    +\beta_{\lambda_d}(\lambda,\lambda_d)
    \frc{\p}{\p\lambda_d}\rt)\,:\Or_j:\; = 0 \quad
    (j=0,1,\ldots,\infty)~.
\eeq
Since the normal ordering operation (\ref{normord}) is also valid
on the Hilbert space ${\cal H}_{-V}$, it is clear that the
expression (\ref{curexp}) also gives the divergent and finite part
of the current operator on ${\cal H}_{-V}$, so that the recursion
relation (\ref{CSrecurs}) among operators $:\Or_j:$ implies the
Callan-Symanzik equation (\ref{CS}) also holds for the
steady-state average of the current. Of course, the same is true
for any operator that has a well-defined average in the steady
state.

All this will be explicitly verified for the current operator to
one loop in the following sub-section.

{\bf Density-matrix-independent calculation of the beta functions
and of the anomalous dimension of the current to one loop.} 
Let us write $\bar\cur_\Lambda$ using time variables instead of momentum
variables:
\beq\label{barcurrt}
    \bar\cur_\Lambda = \sum_{k=0}^\infty i^k \int_{-\infty}^0
    dt_1 \int_{t_1}^0 dt_2 \cdots
    \int_{t_{k-1}}^0 dt_k
    [(H_I^{(0)})_\Lambda(t_1),[(H_I^{(0)})_\Lambda(t_2),
        \cdots,[(H_I^{(0)})_\Lambda(t_k),(\cur^{(0)})_\Lambda(0)]\cdots]]~.
\eeq
Here
\beq
    (H_I^{(0)})_\Lambda(t) = \lt(\lambda (\vec{J}_y)_\Lambda(-t) +
    \lambda_d(\vec{J}_d)_\Lambda(-t)\rt)\cdot\vec{S}
\eeq
and
\beq
    (\cur^{(0)})_\Lambda(t) = \lambda (\vec{J}_x)_\Lambda(-t)\cdot\vec{S}
    ~.
\eeq
In the regularization scheme that we consider, characterized by
the regulator $R_\Lambda(p)$ (\ref{regLambda}), it is a simple
matter to observe that
\beq
    \Lambda\frc{\p}{\p\Lambda} (H_I^{(0)})_\Lambda(t) = -\frc1{\Lambda^2}
    (H_I^{(0)})_\Lambda''(t) ~,\qquad
    \Lambda\frc{\p}{\p\Lambda} (\cur^{(0)})_\Lambda(t) = -\frc1{\Lambda^2}
    (\cur^{(0)})_\Lambda''(t)
\eeq
where primes mean time derivatives. Consider the first few terms
of (\ref{barcurrt}):
\beqa
    \bar\cur_\Lambda &=& (\cur^{(0)})_\Lambda +
    \int_{-\infty}^0dt\,[i(H_I^{(0)})_\Lambda(t),(\cur^{(0)})_\Lambda(0)]
    + \n && +
    \int_{-\infty}^0dt_1\int_{t_1}^0dt_2\,[i(H_I^{(0)})_\Lambda(t_1),
        [i(H_I^{(0)})_\Lambda(t_2),(\cur^{(0)})_\Lambda(0)]]
    + \ldots
\eeqa
Using integration by parts, we find
\beqa
    -\Lambda\frc{\p}{\p\Lambda} \bar\cur &=& \frc1{\Lambda^2}\Bigg(
    (\cur^{(0)})_\Lambda''(0) + \n &&
    +\ [i(H_I^{(0)})_\Lambda'(0),(\cur^{(0)})_\Lambda(0)]
    + \int_{-\infty}^0dt\,[i(H_I^{(0)})_\Lambda(t),
        (\cur^{(0)})_\Lambda''(0)] + \n && +
    \int_{-\infty}^0dt\,[[i(H_I^{(0)})_\Lambda'(t),i(H_I^{(0)})_\Lambda)(t)],
        (\cur^{(0)})_\Lambda(0)] + \n && +
    \int_{-\infty}^0dt\,[(H_I^{(0)})_\Lambda(t),[(H_I^{(0)})_\Lambda'(0),
        (\cur^{(0)})_\Lambda(0)]] + \n && +
    \int_{-\infty}^0dt_1\int_{t_1}^0dt_2\,[i(H_I^{(0)})_\Lambda(t_1),[i(H_I^{(0)})_\Lambda(t_2),
        (\cur^{(0)})_\Lambda''(0)]]
    + \ldots \Bigg)~.
    \label{Lambdader}
\eeqa
We want to evaluate all this at $\Lambda\to\infty$ in order to
find the beta function using the equation (\ref{CSoper}). Inside
the parenthesis, we need only keep the contributions of order
$\Lambda^2\ln^j(\Lambda)$ for non-negative integers $j$. The very
first operator of course does not contribute, but all others do.
These leading contributions have to be compared with the leading
contributions of
\beq\label{betader}
    \lt(\beta_\lambda\frc{\p}{\p\lambda} +
    \beta_{\lambda_d}\frc{\p}{\p\lambda_d}\rt) \bar\cur =
    \lt(\beta_\lambda\frc{\p}{\p\lambda} +
    \beta_{\lambda_d}\frc{\p}{\p\lambda_d}\rt) \lt[(\cur^{(0)})_\Lambda(0) +
    \int_{-\infty}^0dt\,[i(H_I^{(0)})_\Lambda(t),(\cur^{(0)})_\Lambda(0)] +
    \ldots\rt]
\eeq
where we wrote only the terms contributing to the one-loop order.
The beta functions appearing there can be obtained by requiring
that when the derivatives with respect to the couplings are
applied to the operator $i(H_I^{(0)})_\Lambda(t)$, they give the
operator $[i(H_I^{(0)})_\Lambda'(t),i(H_I^{(0)})_\Lambda)(t)]$
(appearing on the third line of (\ref{Lambdader})) in the limit
$\Lambda\to\infty$. The main contribution in this limit of this
operator can be obtained from
\beq\label{commH1H1}
    \frc1{\Lambda^2}[i(H_I^{(0)})_\Lambda'(t),i(H_I^{(0)})_\Lambda)(t)]
    =
    -i\lt((\lambda_d^2+\lambda^2)(\vec{J}_d)_{\sqrt{2}\Lambda}(-t)
    + 2\lambda\lambda_d
    (\vec{J}_y)_{\sqrt{2}\Lambda}(-t)\rt)\cdot\vec{S}
    + \frc1{\Lambda^2} : \Or :
\eeq
where $\Or$ contains products of current algebra operators at the
same point $x=-t$; the explicit form of $\Or$ is not important
here. Equating the leading behavior of this operator as
$\Lambda\to\infty$ with that of
\[
    \lt(\beta_\lambda\frc{\p}{\p\lambda} +
    \beta_{\lambda_d}\frc{\p}{\p\lambda_d}\rt) iH_I^{(0)}(t)
\]
gives
\beq
    \beta_{\lambda_d} = -(\lambda_d^2 +
    \lambda^2) + O(\lambda_d^3,\lambda^4,\lambda_d^2\lambda^2)~,\qquad
    \beta_\lambda = -2\lambda\lambda_d + O(\lambda\lambda_d^2,\lambda^3)~.
\eeq
Note that we had to take the limit $\Lambda\to\infty$ of the
operator $[i(H_I^{(0)})_\Lambda'(t),i(H_I^{(0)})_\Lambda)(t)]$
rather than of the integral where it is involved on the third line
of (\ref{Lambdader}), since the beta function does not depend on
the particular average that we are calculating. The sub-leading
operators in (\ref{commH1H1}) may give contributions to this
integral, but these are two-loop contributions to the anomalous
dimension of the operator $\cur$ (of course, since this operator
should have zero anomalous dimensions, all such contributions
should cancel out).

The total one-loop contributions to the anomalous dimension of
$\cur$ can be verified to be zero by checking that when the
derivatives with respect to the couplings in (\ref{betader}) are
applied to the operator $(\cur^{(0)})_\Lambda(0)$ (the first term
inside the parenthesis), they give the two terms appearing on the
second line of (\ref{Lambdader}) in the limit $\Lambda\to\infty$.
In this limit, the second term on the second line of
(\ref{Lambdader}) can be written
\beq
    \int_{-\infty}^0dt\,[i(H_I^{(0)})_\Lambda(t),\cur''] \sim
    \int_{-\infty}^0dt\,[i(H_I^{(0)})_\Lambda''(t),\cur] =
    [i(H_I^{(0)})_\Lambda'(0),\cur]~.
\eeq
That is, it is equal to the first term. Together, their leading
behavior at $\Lambda\to\infty$ can be obtained from
\beq\label{commH1cur}
    \frc2{\Lambda^2}[i(H_I^{(0)})_\Lambda'(0),(\cur^{(0)})_\Lambda(0)] =
    -2\lambda_d(\cur^{(0)})_{\sqrt{2}\Lambda}(0) + \frc1{\Lambda^2}:\tilde\Or:
\eeq
where $\tilde\Or$ is of the same form as $\Or$. But since
\[
    \lt(\beta_\lambda\frc{\p}{\p\lambda} +
    \beta_{\lambda_d}\frc{\p}{\p\lambda_d}\rt) \cur
    = \lambda^{-1} \beta_\lambda \cur =
    -2\lambda_d \cur + O(\lambda\lambda_d^2,\lambda^3)~,
\]
we immediately conclude that the anomalous dimension of the
current is zero to one loop.

It is a simple matter to verify that the one-loop steady-state
current (\ref{curI}) with the values (\ref{I1}) and (\ref{I2})
indeed satisfies the Callan-Symanzik equation
\beq
    \lt(\Lambda\lt.\frc{\p}{\p \Lambda}\rt|_{\lambda,\lambda_d}
    -2 \lambda\lambda_d \frc{\p}{\p\lambda}
    -(\lambda_d^2 +
    \lambda^2) \frc{\p}{\p\lambda_d}\rt)
    \bra\cur\ket_{ss} \stackrel{\Lambda\to\infty}=
    0~.
\eeq
Note that our derivation of the Callan-Symanzik equation to
one-loop did not use the particular initial density matrix
$e^{-\beta(H_0-VH_z)}$; only the fact that normal-ordered
operators are finite when averaged with this density matrix.

{\bf Two-loop beta functions.} The two-loop result (\ref{curI2})
with two-loop divergent parts (\ref{2loopdiv}) and one-loop finite
and divergent parts (\ref{I1}) and (\ref{I2}) is
\beqa
    \bra\cur\ket_{ss} &\sim& 3\pi V \lambda^2\lt(\frc12 + 2\lambda_d\lt(P(w) +
    \ln\lt(\frc{\Lambda}V\rt)\rt) + \rt.\n && \qquad\quad \lt.
    \lambda_d^2
    \lt[10P(w)+5\ln\lt(\frc{\Lambda}V\rt)-1\rt]\ln\lt(\frc{\Lambda}V\rt)
    +\lambda^2
    \lt[2P(w)+\ln\lt(\frc{\Lambda}V\rt)-1\rt]\ln\lt(\frc{\Lambda}V\rt)
    \rt)~.
\eeqa
This satisfies perturbatively the Callan-Symanzik equation with
beta functions
\beqa
    \beta_{\lambda_d} &=& -\lambda_d^2-\lambda^2 + O(\lambda_d^3,\lambda^2\lambda_d)\n
    \beta_\lambda &=& -2\lambda\lambda_d + \lambda\lambda_d^2 +
    \lambda^3 + O(\lambda\lambda_d^3,\lambda^3\lambda_d)~.
\eeqa
Note that this calculation does not give the third order
coefficient of $\beta_{\lambda_d}$. However, we know the universal
beta functions of the one-channel Kondo model $-(g^2-\frc12g^3)$
and that of the symmetric two-channel Kondo model $-(g^2-g^3)$.
Taking $V=0$ (which does not affect the beta function), these two
cases are obtained respectively at $\lambda=0$, and at
$\lambda_d=\lambda$ upon diagonalization of the matrix of
couplings $J_{\alpha,\alpha'}$.
 These two facts essentially fix the
two-loop beta function to be of the form
\beqa
    \beta_{\lambda_d} &=& -(\lambda_d^2 +
    \lambda^2 - \ld^3 - \lambda^2\lambda_d +\ldots)
    \n
    \beta_\la &=& - (2\lambda\lambda_d - a\la\ld^2 - (2-a)\la^3 + \ldots)
\eeqa
where $a$ is a non-universal number. Our results fix $a=1$, which
gives the standard beta functions
\beqa
    \beta_{\lambda_d} &=& -(\lambda_d^2 +
    \lambda^2 - \ld^3 - \lambda^2\lambda_d +
    O(\lambda_d^4,\lambda^2\lambda_d^2,\lambda^4))
    \n
    \beta_\la &=& - (2\lambda\lambda_d - \la\ld^2 - \la^3 +
    O(\lambda\lambda_d^3,\lambda^3\lambda_d)) ~.
    \label{betafct}
\eeqa

It is convenient now to change variables to $\lambda_\pm =
\lambda_d\pm\lambda$ so that (to the same order)
 \beq \beta_\pm =
-\lambda_\pm^2 + \frc12 \lambda_\pm(\lambda_+^2+\lambda_-^2)~.
\eeq
The RG invariant anisotropy parameter can be expressed as:
 \beq\label{RGinv}
 C =\frc12\lt(\frc1{\lambda_-}\lt(1-\frc12\lambda_+\rt)
 -\frc1{\lambda_+}\lt(1-\frc12\lambda_-\rt)\rt)
\eeq

{\bf The scaling limit of the current.} We now evaluate the
current as function of the voltage and of the temperature. For the
expression of the current, our analysis will make use solely of
the one-loop results (\ref{curoneloop}).

Expression (\ref{curoneloop}) is the perturbative expansion in the
limit $T,V\ll\Lambda$ of the model. The couplings
$\lambda_d,\,\lambda$ take their ``microscopic'' values: the
actual values for the physical process represented by the
interaction terms, and the cut-off $\Lambda$ is of the scale of
the band width. Of course, since we neglected terms vanishing as
$\Lambda\to\infty$, this is not an exact expression for finite
$\Lambda$. However, it allows us to have an exact expression in
the {\em scaling} limit.

As explained earlier, we expect universality for $T, V \ll
\Lambda$. It is convenient therefore to consider the limit where
$\Lambda$ is sent to infinity with the couplings becoming cut-off
dependent: they are modified so as to keep the physics unchanged
when the cut-off is increased. Denoting for the moment $\Lambda =
\Lambda_0$ the physical value of the cut-off (of the order of the
band width) and $\lambda= \lambda^0, \lambda_d=\lambda^0_d$ the
values of the coupling there, the running couplings
$\lambda^r(\Lambda), \lambda^r_d(\Lambda)$, are governed by the RG
equations,
\beqa\label{RGflow1}
\Lambda \frc{d\lambda_d^r}{d\Lambda} &=&
   (\lambda_d^r)^2+(\lambda^r)^2-(\ld^r)^3 -
   (\lambda^r)^2\lambda_d^r\n
\Lambda\frc{d\lambda^r}{d\Lambda} &=& 2
   \lambda^r\lambda_d^r - \la^r(\ld^r)^2 - (\la^r)^3
\eeqa
 with
   initial conditions fixed by the microscopic values of the
   couplings: $\lambda_d^r(\Lambda=\Lambda_0) =
   \lambda^0_d,\,\lambda^r(\Lambda=\Lambda_0) = \lambda^0$.

The solution of the RG flow can then be described by RG invariants
- quantities that describe the full trajectory. Such quantities
are $C$ introduced earlier, and a scale $T_K$ to be discussed
below. Thus any physical quantity $F$ will depend on the cut off
and coupling via these invariants $F = F(T/T_K, V/ T_K, C)$. One
may reformulate this scaling procedure as follows. Introduce the
scale $M$, the physical scale on which the system is examined,
\[
    V=M\sin(\alpha)~,\quad T=M\cos(\alpha)
\]
for some angle $\alpha$ in the $V-T$ plane so that the previous
result for  the current is written as,
\[
 \bra\cur\ket_{ss} =
\frc{3\pi}2 \, V\, \lambda^2\lt[1+4 \lambda_d \;
\lt(P(\tan(\alpha)) + \ln\lt(\frc{\Lambda}M\,\csc(\alpha)\rt)\rt)
+ \ldots\rt]
\]
Following the previous considerations, the current can be written
in terms of couplings that depend on the physical scale $M$,
satisfying the equations with respect to $M$,
\beqa\label{RGflow}
M \frc{d\lambda_d^r}{dM} &=&
   -(\lambda_d^r)^2-(\lambda^r)^2+(\ld^r)^3 +
   (\lambda^r)^2\lambda_d^r\n
M\frc{d\lambda^r}{dM} &=& -2
   \lambda^r\lambda_d^r + \la^r(\ld^r)^2 + (\la^r)^3
\eeqa
with initial conditions
$\lambda_d^r(M=\Lambda)=\lambda_d,\,\lambda^r(M=\Lambda)=\lambda$
($\Lambda$ being of the order of the band width), as follows:
 \beq\label{curRGimproved}
\bra\cur\ket_{ss} = \frc{3\pi}2 \, V\, (\lambda^r)^2\lt[1+4
\lambda_d^rQ(\alpha)+ \ldots\rt]
\eeq
 where
\beq\label{Q}
    Q(\alpha) = P(\tan(\alpha)) + \ln(\csc(\alpha))~.
\eeq

Again, the solution to the RG flow (\ref{RGflow}) should be
described solely as a function of $M/T_K$ and of the RG invariant
$C$ (\ref{RGinv}), instead of $\Lambda$ and the initial conditions
$\lambda_d$ and $\lambda$. With such a description, we can
trivially take the scaling limit: $T,\,V,\,T_K\ll\Lambda$ with
fixed ratios $T : V : T_K$ and fixed $C$, since $\Lambda$ does not
appear anymore. In this limit, the quantum field theory gives
exact results, and the system is in its universal regime. In order
to have unambiguous results, we need to define $T_K$. Once $T_K$
is defined, one need only solve the RG flow (\ref{RGflow}) (a
numerical solution is easy to obtain with good precision, for
instance), and one obtains the scaling limit in its perturbative
region. The actual values of $\lambda_d,\,\lambda$ and of the band
width $\Lambda$ should be such that the theory is near to the
scaling limit if we want the system to be meaningfully described
by quantum field theory.

The perturbative region of the scaling limit is $T_K \ll
\sqrt{V^2+T^2}$, for any value of $C$. Hence, in order to define
$T_K$ in perturbation theory, we need to look at the expansion as
$M/T_K \to\infty$ of the solution to the RG flow (\ref{RGflow}).
This expansion, and the value of $T_K$, have different forms
depending on $C$. We will consider here two cases: taking
$C=\infty$ then $M/T_K\to\infty$; and keeping $C$ finite then
taking $M/T_K\to\infty$. The first case corresponds to
$\lambda_d=\lambda$. In fact, it is true to all orders that if
$\lambda_d=\lambda$, then $\lambda_d^r=\lambda^r$ for all scales
$M$. In this case, the equilibrium Hamiltonian $H|_{V=0}$ is in
fact a one-channel Kondo model plus a decoupled free massless
fermion. The case $C$ finite corresponds naturally to $\lambda\sim
C\lambda_d^2$ in the scaling limit (as will become clear below).

In the case where $\lambda_d=\lambda$, the RG equations lead to
the following large-($M/T_K$) expansion of the running couplings:
\beq\label{largeMeq}
    \ld^r = \la^r = \frc1{2\ln(M/T_K)} +
    \frc{\ln\ln(M/T_K)}{4\ln^2(M/T_K)} + \frc{a}{2\ln^2(M/T_K)}
    + \frc{\ln^2\ln(M/T_K)}{8\ln^3(M/T_K)} +
    \frc{(4a-1)\ln\ln(M/T_K)}{8\ln^3(M/T_K)} +
    O\lt(\frc1{\ln^3(M/T_K)}\rt)~.
\eeq
The coefficient $a$ can be changed by a change of the scale (the
integration constant) $T_K$; making the replacement $T_K \mapsto
xT_K$ is equivalent to doing $a\mapsto a+\ln(x)$. Note also that a
change of scale $T_K \mapsto xT_K$ corresponds simply to a
perturbative change of the running coupling constants that keeps
the beta functions invariant. Fixing the value of $a$ is making a
choice of definition for $T_K$, which is one more condition
necessary to totally specify the renormalization procedure. Of
course, different definitions of $T_K$ reproduce the same scaling
limit. For arbitrary $a$, in terms of the coupling
$\lambda_d=\lambda$ and of the scale of the band width $\Lambda$,
$T_K$ has the form
\[
    T_K(a) = \Lambda\,\sqrt{2
    \lambda}e^{-a-\frc1{2\lambda}}(1+O(\lambda))\qquad (\ld=\la)~.
\]
A standard definition for the Kondo temperature is to make the
term in $1/\ln^2(M/T_K)$ vanish ($a=0$), giving:
\beq\label{TKondo1}
    T_K = \Lambda\,\sqrt{2
    \lambda}e^{-\frc1{2\lambda}}(1+O(\lambda))\qquad (\ld=\la)~.
\eeq

In the case where the RG invariant $C$ is finite, the large-($M/T_K$)
expansion has the form
\beqa
    \ld^r &=& \frc1{\ln(M/T_K)} +
    \frc{\ln\ln(M/T_K)}{\ln^2(M/T_K)} + \frc{a}{\ln^2(M/T_K)}
    + \frc{\ln^2\ln(M/T_K)}{\ln^3(M/T_K)} +
    \frc{(2a-1)\ln\ln(M/T_K)}{\ln^3(M/T_K)} +
    \ldots \n
    \frc{\la^r}C &=& \frc1{\ln^2(M/T_K)} +
    \frc{2\ln\ln(M/T_K)}{\ln^3(M/T_K)} + \frc{1+2a}{\ln^3(M/T_K)}
    + \frc{3\ln^2\ln(M/T_K)}{\ln^4(M/T_K)} +
    \frc{(6a+1)\ln\ln(M/T_K)}{\ln^4(M/T_K)} +
    \ldots \n
    \label{largeM}
\eeqa
where the dots ($\ldots$) mean $O(\ln^{-3}(M/T_K))$ for
$\lambda_d^r$, and $O(\ln^{-4}(M/T_K))$ for $\lambda^r$. Note that
in general, $C$ will not appear only as a normalization of $\la$;
this is an artifact of the limited perturbative order which we
consider. Also, note that $C$ must be positive for $\lambda^r$ to
be positive. Again, a variation of $T_K$ has the effect of
changing $a$: making the replacement $T_K \mapsto xT_K$ is
equivalent to doing $a\mapsto a+\ln(x)$. For arbitrary $a$, we can
use the first or the second equation of (\ref{largeM}) in order to
determine $T_K$ in terms of the couplings $\lambda_d,\,\lambda$
and of the scale of the band width $\Lambda$. This gives two
equivalent forms:
\[
    T_K(a) =
    \Lambda\,\lambda_de^{-a-\frc1{\lambda_d}}(1+O(\lambda_d))=
    \Lambda\,\sqrt{\frc{\lambda}{C}}e^{-a-\frc12-\sqrt{\frc{C}{\lambda}}}(1+O(\sqrt{\lambda}))
     \qquad
    (\lambda_d\neq\lambda)~.
\]
Recall that from (\ref{RGinv}), we indeed have that
$\lambda\propto\lambda_d^2$ in the scaling limit, so that the
second equality above is correct. A more standard way of writing
the Kondo temperature can be obtained by considering the linear
combination $\lambda_d^r + \lambda^r$. This gives, again for the
same object $T_K(a)$,
\[
    T_K(a) =
    \Lambda\,(\lambda_d+\lambda)e^{-(a+C)-\frc1{\lambda_d+\lambda}}(1+O(\lambda_d))
     \qquad
    (\lambda_d\neq\lambda)~.
\]
A standard definition is $a=-C$, giving
\beq\label{TKondo2}
    T_K =
    \Lambda\,(\lambda_d+\lambda)e^{-\frc1{\lambda_d+\lambda}}(1+O(\lambda_d))
     \qquad
    (\lambda_d\neq\lambda)~.
\eeq

Four comments are now in order.

First, note that taking $C\to\infty$ in (\ref{largeM}) does not give (\ref{largeMeq}).
This is expected, since the former corresponds to taking first $M/T_K\to\infty$ then
$C\to\infty$ in the solution to the RG equations,
whereas the latter corresponds to taking first $C\to\infty$ then $M/T_K\to\infty$,
and these two limits do not commute. In particular, one would have obtained still different
expressions taking simultaneously $C\to\infty$ and $M/T_K\to\infty$, with a prescribed ratio
between them, for instance (it is a simple matter to evaluate in this case, or in any other case,
the expansion as $M/T_K\to\infty$ of the solutions to the RG equations).
The expressions (\ref{largeM}) and (\ref{largeMeq}) should
not be regarded as describing all solutions to the RG equation; they are rather particular limits of
the solutions, as described above, and are used here solely for the purpose of defining the Kondo
temperature purely using perturbation theory (the perturbative regime is, again, the
regime $M\gg T_K$, for any $C$ taken to infinity or not, simultaneously or not with $M/T_K$).
The expansion (\ref{largeM}) offers a way of defining $T_K$ in the
regime with $C$ finite, whereas the expansion (\ref{largeMeq}) offers a way of defining it
in the regime with $C=\infty$. Other definitions would have been possible, but we 
only need these two definitions of the Kondo temperature here.

Second, it is important to recall that the cases $C=\infty$ and $C<\infty$
do exhaust all possible scaling regimes, even though the expressions
(\ref{largeMeq}) and (\ref{largeM}) do not exhaust all possible behavior of the
running couplings as $M/T_K\to\infty$. However, there are many ways
of reaching any given scaling regime. In particular, the relations $\lambda=\lambda_d$ and
$\lambda\sim C\lambda_d^2$ (naturally associated to, respectively, the scale definitions
(\ref{TKondo1}) and (\ref{TKondo2})) do not exhaust all possible
ways the same scaling limit can be reached. In order to understand
what this means, first recall that the full
RG-improved perturbation theory reproduces the divergent and
finite part of the full bare perturbation theory, so that we can talk
about the RG trajectory for the bare couplings before
taking the scaling limit. Then, from this viewpoint, one can take the scaling limit by fixing a
trajectory $C$, and by sending $\Lambda\to\infty$ while keeping
$\lambda_d$ and $\lambda$ on the trajectory at scale $\Lambda$. For finite $C$,
this gives $\lambda\sim C\lambda_d^2$, whereas for $C=\infty$, this gives $\lambda=\lambda_d$. 
But one could also take the scaling limit by changing the value of $C$
(changing the shape of the trajectory) while $\Lambda\to\infty$,
always keeping $\lambda_d$ and $\lambda$ on the trajectory defined
by $C$. For instance, this is what happens if one takes
$\lambda=q\lambda_d$ for some fixed $q\neq1$ when sending
$\lambda_d,\lambda\,\to0$; then one must simultaneously take $C\to\infty$
in order to keep $\lambda$ and $\lambda_d$ on the RG trajectory.
The resulting value of $T_K$ in terms of such couplings is different from (\ref{TKondo1}) and
(\ref{TKondo2}). But in this example, when the scaling limit is
reached we have $C=\infty$, so that the quantum field theory is the same as the one obtained by taking
$\lambda=\lambda_d$ and sending them to 0; in particular, in the scaling limit, we still
have $\lambda^r=\lambda_d^r$, and we can still define a scale $T_K$ using (\ref{largeMeq}).
Since we are only interested in the scaling regimes, we do not need to look at all possible
ways a given regime can be reached.

Third, it is important to note that the leading behavior of the current
at large $M/T_K$ is very different if $C=\infty$ or if $C<\infty$ (again, we only look at the two
cases mentionned above). In the first case, it is given by
\beq
    \bra\cur\ket_{ss} \sim
    \frc{3\pi }{8}\,\frc{M\sin(\alpha)}{\ln^2(M/T_K)} = \frc{3\pi }{8}\,
    \frc{V}{\ln^2(\sqrt{V^2 +T^2}/T_K)}\quad
    (\lambda_d=\lambda)~.
\eeq
On the other hand, in the second case it is given by
\beq
    \bra\cur\ket_{ss} \sim
    \frc{3\pi C^2}{2}\,\frc{M\sin(\alpha)}{\ln^4(M/T_K)} = \frc{3\pi C^2}{2}\
    \frc{V}{\ln^4(\sqrt{V^2 +T^2}/T_K)}
 \quad
    (\lambda_d\neq\lambda)~.
\eeq
Note that since the RG invariant $C$ appears as a coefficient, the
leading behavior in this case is non-universal; this is a property of the regime $C$ finite.

Finally, the parameter $\alpha$ can be seen as parametrizing a family of
choices of infrared cutoffs for our theory. For instance, at
$\alpha=0$, the temperature is the infrared cutoff, whereas at
$\alpha=\pi/2$, the voltage solely plays the role of an infrared
cutoff. It is important to note that the voltage is a good
infrared cutoff for the average of the current operator to
two-loop order. Indeed, the limit $T/V\to0$ of our one-loop and
two-loop bare perturbative results is finite; equivalently, the
limit $\alpha\to\pi/2$ of the renormalized perturbative results is
finite. We should remark however that this need not be the case
when other quanties are considered, see \cite{rg1}.

{\bf Universal ratios.} From the viewpoint of the interpretation
of the measurement of non-equilibrium quantities, the Kondo
temperature, as defined for instance in (\ref{TKondo1}), is not
very convenient, since
 the temperature $T$ and the bias voltage $V$ have a
different influence on the Kondo screening cloud, and we would
like a definition that embodies this difference. Hence, it seems
appropriate to define a continuum of scales, $M_K(\alpha)$,
depending on the angle $\alpha$ on the $V-T$ plane.
  Consider first the case $\lambda_d=\lambda$. A possible definition
of $M_K(\alpha)$ is the requirement that the average current
$\cur$ in the steady state does not have a term of the form
$M/\ln^3(M/M_K(\alpha))$ in its expansion at large
$M/M_K(\alpha)$:
\beq
    \bra\cur\ket =
    \frc{3\pi }{8}\,\frc{M\sin(\alpha)}{\ln^2(M/M_K(\alpha))}\lt(
    1+ \frc{\ln\ln(M/M_K(\alpha))}{\ln(M/M_K(\alpha))} +
    O\lt(\frc{\ln^2\ln(M/M_K(\alpha))}{\ln^4(M/M_K(\alpha))}\rt)\rt)~.
\eeq
This gives:
\beq\label{MK}
    M_K(\alpha) = T_K \,e^{Q(\alpha)}
\eeq
where $T_K$ is defined in (\ref{TKondo1}).

 As particular cases, we can now define the ``decoherence'' Kondo
scale, $M_K(\pi/2)$, as the voltage at $T=0$ at which the Kondo
cloud is destroyed by the electrons passing through, and the
``thermal'' Kondo scale $M_K(0)$. We find that their ratio, which
is universal and does not receive corrections from the two-loop
(or higher-loop) contributions to the beta function (nor from
higher order contributions to the current), is
\beq R_{\mbox{current}} = \frc{M_K(\pi/2)}{M_K(0)} = 2\pi
e^{1-\frc{3\gamma}2 - \frc{\sqrt{\pi}}2} = 2.96188723...  \eeq In
the case $\lambda_d\neq\lambda$ (that is, for $C<\infty$), a
similar definition of $M_K(\alpha)$ can be made: requiring that
the current does not have a term of the form
$M/\ln^4(M/M_K(\alpha))$. Constructing again the ratio
$M_K(\pi/2)/M_K(0)$ gives exactly the same number: this ratio is
indeed universal, independent of both $T_K$ and $C$. It is
important that the current possess an infrared-convergent
(convergence at large switch-on times) 
perturbative expansion to one loop for these universal ratios to
have a meaning. Further, that they have the usual one-loop
logarithmic accuracy near to the scaling limit is a consequence of
the infrared convergence of the two-loop perturbation theory.

The scale $M_K(\alpha)$ defined above is characteristic of the
current; other physical quantities would give different functions
$M_K(\alpha)$, and different ratios. For instance, the same
analysis can be applied on the differential conductance
\[
    G = \frc{d}{dV} \bra\cur\ket_{ss}~.
\]
From the perturbative calculations, we have
\beq
    G = \frc{3\pi}2\, \lambda^2\lt[1+4
    \lambda_d\lt(P\lt(\frc{V}T\rt)+\ln\lt(\frc{\Lambda}{V}\rt) + \frc{V}T
    P'\lt(\frc{V}T\rt) - 1 \rt) + \ldots\rt]~.
\eeq
Again, in terms of running couplings, we have
\beq
    G = \frc{3\pi}2\, (\lambda^r)^2\lt[1+4
    \lambda_d^r\tilde{Q}(\alpha) + \ldots\rt]
\eeq
where
\beq
    \tilde{Q}(\alpha) = P(\tan(\alpha))+\ln(\csc(\alpha)) +
    \tan(\alpha)
    P'(\tan(\alpha)) - 1~.
\eeq
Taking again $\lambda_d=\lambda$, we can repeat the calculations
above and define similarly the scale $\tilde{M}_K(\alpha)$
associated to the conductance. We find
\beq
    \tilde{M}_K(\alpha) =
        T_K\,e^{\tilde{Q}(\alpha)}~.
\eeq
We observe that $\tilde{Q}(0) = Q(0)$ and that $\tilde{Q}(\pi/2) =
Q(\pi/2) - 1$. Hence, we have
\beq
    R_{\mbox{conductance}} =
    \frc{\tilde{M}_K(\pi/2)}{\tilde{M}_K(0)} = e^{-1}
    R_{\mbox{current}} = 2\pi e^{-\frc{3\gamma}2 -
    \frc{\sqrt{\pi}}2}  = 1.08961742...
\eeq

\section{Perspectives}

{\bf Reaching the steady-state.} We showed that the bath of free
massless fermions suffices to allow the system to reach
equilibrium at temperature $T$ in the case of zero bias voltage,
and to allow it to reach steady state when the bias voltage is
non-zero. No other relaxation process has to be assumed; the
infinite bath of free massless fermions plays the role of a
thermal bath and is able to absorb the energy necessary for
relaxation to occur, as well as for the steady state to exist.
This is in close connection with the study of Caldeira and Leggett
\cite{CaldeiraL81}: they constructed a model where an infinite
number of oscillators provides an explicit dissipation in order to study,
from first principles, the effect of dissipation on quantum
tunneling. It turns out \cite{CallanT90} that their construction
is simply related to models of field theory that are conformal in
the bulk (in general, with non-conformal boundary conditions), as recall
in the Introduction. It
would be interesting to see to what extent our proof can be
generalized to the study of more general models of quantum
kinetics with thermal dissipation, and to see whether
similar arguments can be applied to other impurity models out of equilibrium.
A principle to follow, as can be extracted from our derivation, is the fact that
the orbit of a global symmetry of the model should cover the possible values of the
boundary degrees of freedom.

Related to the latter point, another question is: What is the effect of a magnetic field on the
steady state in the quantum dot? Our proof of factorization in
Appendix \ref{factorization} does not hold anymore when a magnetic field is present,
since it uses heavily the invariance of the correlation functions
under $SU(2)$ transformations. Hence, the real-time perturbation
series is no longer expected to be convergent as the switch-on time is sent to minus infinity.
This is simple to understand physically: at small couplings, both the current through the dot
and its interaction with the leads are weak, that is, both the non-equilibrating effect
and the thermalization effect are weak. At zero magnetic field, since all states of the isolated dot have the
same energy, the thermalization effect is more efficient and still stronger. At non-zero magnetic
field, however, as the couplings are sent to zero, we cannot
expect that the dot smoothly reaches its thermal equilibrium energy distribution when no other thermal bath
is coupled to it.
The real-time perturbation theory should describe this situation, but obviously its
zeroth order cannot give anything else than the thermal equilibrium value of any quantity under
study. Hence, in non-zero magnetic field, we can expect some strong non-analyticity in the couplings and
we must find large-time (IR) divergences in the perturbative coefficients.
This clearly indicates that the real-time perturbative series does not properly describe the
approach to the steady state, neither the steady state itself, of the model in magnetic field without
external thermal bath.

Questions remain, as raised in the Introduction:
Does the quantum field theory still (non-perturbatively) reach a steady state, or
does it show other behaviors at large times, like oscillations (of
the dot magnetization, for instance)? If it reaches a steady state, is it a good description of
realistic systems, where the dot is coupled to an external thermal bath
at all times (without exchange of particles), independently of its coupling to the leads?

The former question was partially answered, in \cite{rg6}: it was assumed on physical grounds that the
non-equilibrium Kondo model reaches a steady state, and mainly
from this assumption, it was explained how to obtain the zeroth
order of perturbation theory for the dot magnetization. Indeed, strong non-analyticity is obtained. In a
sense, one should start with a density matrix that already
contains a non-thermal distribution of the impurity spin states;
this density matrix can be obtained by requiring that the
perturbative series be convergent at large times. As explained clearly there, this is equivalent to
solving a quantum Boltzmann equation in order to determine the non-thermal dot occupation numbers.
This should answer partly, in some sense still perturbatively,
the question of describing the steady state of the quantum field theory (the non-equilibrium Kondo model
with magnetic field) -- although the results really start with the assumption of a steady state, and do not
establish its existence. However, this does not address the question as to whether the leads correctly play the role
of thermal baths, or whether a coupling to an external thermal bath would have important effects.

In relation to the latter question, our argument suggests that, at least at small couplings (or
at temperatures much greater than the Kondo temperature), the model
{\em does not} describe the true steady state of the non-equilibrium Kondo dot in contact
with a thermal environment. Indeed, from
a physical interpretation of the perturbative series, the leads do not provide
a strong enough thermalisation to absorb the energy necessary for the steady state to occur,
and it is possible that it cannot be trusted to
sustain the correct steady state. That it does not provide the thermalisation
for the steady state to occur is certainly in agreement with \cite{rg6}: there it was one of the main points that
the large-time divergences are due to the absence of a proper thermal bath, and that one needs to put
``by hand'' a thermal bath connected to the dot. This was done,
essentially, by putting a small imaginary
part on the evolution time, which was then set to zero before taking small couplings $\lambda,\lambda_d$
(that is, it was obtained a steady state where the coupling to the thermal bath is much smaller than the couplings
$\lambda,\lambda_d$).
The most delicate question, however, concerns the fact that the steady state itself may
be affected non-trivially by a thermal bath. It is probable (but this should be verified) that other, more realistic,
representations of a thermal bath give the same results as \cite{rg6}
in the limit of small coupling with the thermal bath. However,
as the couplings $\lambda,\lambda_d$ are sent to zero, due to the thermal effect of the environment,
the average dot magnetization, for instance, should smoothly reach its thermal equilibrium
value, and this may well be a universal cross-over behavior (since it occurs at small couplings, hence near to a
second order phase transition). In order to assess this, it would be important to have a more adequate
description of a thermal bath, and the theory of Caldeira and Leggett surely provides the most promising avenue.

{\bf The steady-state density matrix.} Our proof that the
real-time perturbation theory describes a steady state was
immediately adapted to the proof that the steady state can be
described by a density matrix (\ref{opY}), almost as in an
equilibrium state. As mentioned, a steady-state density matrix was
also introduced in \cite{Hershfield93} for generic models under
the assumption that a relaxation time was present. Our derivation
is slightly different, and does not make further assumptions,
since we showed that relaxation does occur in the non-equilibrium
Kondo model. The main characteristics of the steady-state density
matrix, as opposed to equilibrium density matrices, is that it is
defined by coupling a {\em non-local} conserved charge $Y$ to the
voltage $V$, instead of coupling a local conserved charge to an
appropriate chemical potential. This non-locality is related to
the fact that the operator $Y$ describes the build-up of the
steady state, or, in some sense, the asymptotic state that
characterizes the steady state. The properties of this asymptotic
states and other consequences of this description are still to be
explored.

{\bf Renormalized real-time perturbation theory.} We
developed the two-loop renormalized real-time perturbation theory.
We gave an argument for the validity of the Callan-Symanzik
equation in the steady state (with the same beta functions as the
equilibrium ones), and verified this to one loop for the current
operator inside any correlation function. It is tempting to relate
the validity of the Callan-Symanzik equation to the fact that the
operator $Y$ appearing in the steady-state density matrix is a
conserved charge. In particular, the fact that the voltage does
not flow is obvious from such consideration, as it is coupled to a
conserved charge. However, since $Y$ is non-local, it is hard to
make this connection more precise.

The quantum field theory gives physical quantities in the
scaling limit (the universal region) $V,\,T,\,T_K \ll
D$, where $T_K$ is the Kondo temperature, which is a non-universal
quantity related to the microscopic values of the couplings.
Our renormalized perturbative results give the current in the region
$T_K \ll \sqrt{V^2 + T^2} \ll D$. This includes the part of the
universal region where the system is strongly out of equilibrium.
In this perturbative region, we defined a continuous family of Kondo scales
$M(\alpha)$ depending on the ratio $V/T = \tan(\alpha)$. By
comparing the scale at $\alpha = \pi/2$ ($T/V\to0$) with that at
$\alpha=0$ ($V/T\to0$), we obtained a universal measure of the
effect of the voltage on the Kondo cloud, as compared to the
effect of the temperature. We noted that such a universal measure
was correct since to two-loop order, the voltage plays the role of
a good infrared cutoff for the current, so that no
divergencies appear to that order as $T/T_K\to0$ (with fixed $V/T_K$).

In connection to the latter point, it has sometimes been suggested in the
literature that at $T/T_K=0$, the system should be in a ``strong
coupling regime'' (see for instance \cite{ColemanHP}), and as such, the perturbation theory should not
be valid and should show infrared divergencies (divergencies as
$T/T_K\to0$) in higher-loop calculations. In particular, it is clear
that this occurs in the calculation of any thermodynamical quantities, which can
indeed be deemed ``in a strong coupling regime'' (at the IR fixed point) at zero temperature.
In \cite{rg2,rg3}, for instance, one sees logarithmic divergences as $T/T_K\to0$ for any fixed $V/T_K$
at the one-loop order of perturbation theory for the spin susceptibility, pointing to the
fact that for describing the limit $T/T_K\to0$ of that thermodynamical quantity,
one needs to know about the IR fixed point\footnote{The large-$V/T_K$ behavior in $1/V$ found at 0-loop order,
then, can only be trusted in the region $V\gg T \gg T_K$.}
(the perturbation theory only describes the theory around its UV fixed point). Our point, though, is that
this might not be so for all quantities. Our two-loop results suggest that in a sense,
the current is really a dynamical quantity, ruled by the scale $V$. For comparison, this
is much like a correlation function of two local fields is ruled by the distance between the points
in equilibrium quantum field theory, even at zero temperature (the short distance behavior
is described, for instance, by the UV fixed point of the theory, up to a normalization
if the fields are not conserved currents, and up to the one-point functions of the operators appearing
in the operator product expansion of the fields). Then, the region
$V\gg T_K$ should really be, for the current, a weak-coupling, UV situation, even
at zero temperature; this is at least what we see at two-loop order. Notwithstanding the fact that infrared
divergencies may signal that the steady state is {\em not}
reached, as discussed above, possible divergencies in the perturbative expansion
of the current at higher orders may correspond to
simple power-like non-analyticity in the couplings at zero temperature, with smaller
contributions than those of the two-loop results (as in the usual situation of correlation
functions in equilibrium quantum field theory).

Finally, it would be very interesting to fully verify the validity of the
Callan-Symanzik equation with a magnetic field. It is easy to check for instance
that the one-loop, finite-magnetic-field results of \cite{rg2, rg3, rg5}
satisfy the one-loop Callan-Symanzik equations in the universal regime. Although we did not cover
the case with a finite magnetic field (and see the discussion above
for the subtleties involved), we expect that our general arguments
for the validity of the Callan-Symanzik equations to all order
still hold since in the universal regime, the magnetic field is a
low-energy scale as compared to the band width. It may be useful to
note that the usual perturbative renormalization 
was modified in \cite{rg2,rg3} in order to correctly incorporate
the structure of logarithmic divergencies in the region $V>D$
(where $D$ is the bandwidth) of the one-loop calculation of the current and of the
impurity magnetization at finite magnetic field; more precisely, energy-dependent
coupling constants were introduced.
We want to stress that this region may be non-universal (that is,
results may depend on the precise structure of the band), as is the region $T>D$ or $T_K>D$.
Then, naturally, it cannot be universally described by
a finite number of coupling constants (or, more precisely, by a finite
number of RG invariants). In order to recover ``scaling'', one needs to use
exact RG or similar methods, and the usual Callan-Symanzik equation does not hold;
but this is not in disagreement with our results, which only deal with the universal regime
$V,T,T_K \ll D$.

\bigskip

{\bf Acknowledgments}

BD is grateful to J.  Cardy, J. Chalker and F. Essler, as well as all
members of the QFT research group at Oxford, for sharing their
insights in many occasions, to A. Lamacraft for discussions during
his visit, and to N. Shah for useful comments on the manuscript.
BD acknowledges support from an EPSRC post-doctoral fellowship (grant GR/S91086/01),
and also acknowledges Rutgers University, where this work was
initiated. NA is grateful to C. Bolech, P. Mehta, O. Parcollet, A.
Rosch and A. Schiller, for numerous illuminating discussions,
criticisms and suggestions as well as useful comments on the
manuscript. We would like to thank S. Kehrein for various
discussions during the early stage of this work.

\appendix

\section{Connected correlation functions}

\label{connected}

Consider a theory with density matrix $\rho$, and denote, for any
operator $\Or$,
\[
    \bra\bra\Or\ket\ket =
    \frc{\Tr\lt[\rho\Or\rt]}{\Tr\lt[\rho\rt]}~.
\]
Consider the following average:
\beq
    \bra\Or\ket \equiv \frc{\lt\bra\lt\bra \Pexp\lt( i\int_{t_0}^{0} dt\,H_I(t)\rt)
    \Or\rt\ket\rt\ket}{
    \lt\bra\lt\bra \Pexp\lt( i\int_{t_0}^{0} dt\,H_I(t)\rt)
    \rt\ket\rt\ket}
\eeq
where $H_I(t)$ can be any operator depending on the (time)
parameter $t$, and $\Or$ is also any operator. The path-ordered
exponential above is understood as an expansion in time-ordered
integrals of multilinears of $H_I(t)$'s. The connected correlation
functions, where the $H_I(t)$'s are ``connected'' to $\Or$, can
then naturally be defined by saying that
\beq
    \bra\Or\ket = \sum_{n=0}^\infty i^n
    \int_{t_0}^0 dt_1 \int_{t_1}^0dt_2 \cdots \int_{t_{n-1}}^0
    dt_n \, \bra\bra H_I(t_1)H_I(t_2)\cdots H_I(t_n)\,\Or\ket\ket_{\rm
    connected}~.
\eeq
Equivalently, they can be defined recursively by
\beqa
    \label{defconnected}
&&    \bra\bra H_I(t_1)\cdots H_I(t_n) \,\Or
    \ket\ket \n
&&  \qquad =
    \bra\bra H_I(t_1)\cdots H_I(t_n) \,\Or
    \ket\ket_{\rm connected} \\
&&  \qquad\quad+
    \sum_{m=1}^n\hspace{-1cm}
    \sum_{\tiny\begin{array}{c}\{\alpha_1,\ldots,\alpha_m\}\subseteq\{1,\ldots,n\}\\
        \alpha_1<\cdots<\alpha_m\\
        \{\beta_1,\ldots,\beta_{n-m}\}\cup\{\alpha_1,\ldots,\alpha_m\}=\{1,\ldots,n\}
        \end{array} }\hspace{-2cm}
    \bra\bra H_I(t_{\alpha_1})\cdots H_I(t_{\alpha_m})\ket\ket\;
    \bra\bra H_I(t_{\beta_1})\cdots H_I(t_{\beta_{n-m}}) \,\Or
    \ket\ket_{\rm connected}~. \no
\eeqa
This recursive definition does not involve integrations over the
time parameters $t_i$'s, so that it is in fact a general
definition for correlation functions where a set of arbitrary
operators $H_I(t_i)$'s parametrized by $i=1,\ldots,n$ are
connected to $\Or$.

The main property of connected correlation functions is the
following. Consider a fixed set of $t_i$'s for $i=1,\ldots,n$ and
a fixed subset $t_{\alpha_1},\ldots,t_{\alpha_m}$ for $m<n$.
Consider correlation functions of $H_I(t_i)$'s with and without
insertion of $\Or$. One can observe that if the operators
$H_I(t_{\alpha_i})$'s factorize out of all such correlation
functions (for instance:
\beqa
&&    \bra\bra H_I(t_1)\cdots H_I(t_{\alpha_1})\cdots
    H_I(t_{\alpha_m})\cdots
    H_I(t_n)\,\Or\ket\ket \n
&&  \qquad\qquad
    = \bra\bra
    H_I(t_1)\cdots \widehat{H_I(t_{\alpha_1})}\cdots \widehat{H_I(t_{\alpha_m})}\cdots
    H_I(t_n)\,\Or\ket\ket \; \bra\bra
    H_I(t_{\alpha_1})\cdots H_I(t_{\alpha_m})\ket\ket\no~)
\eeqa
then all connected correlation functions involving at least one of
the $H_I(t_{\alpha_i})$'s are zero. This is easy to show from
(\ref{defconnected}) by induction on the number of operators
$H_I(t_i)$'s inside connected correlation functions.

Connected correlation functions also occur in more general
situations:
\beqa
&&    \frc{\lt\bra\lt\bra \Pexp\lt( i\int_{t_1^i}^{t_1^f}
        dt\,H_I^{(1)}(t)\rt)
    \ \Pexp\lt( i\int_{t_2^i}^{t_2^f} dt\,H_I^{(2)}(t)\rt) \cdots
    \Pexp\lt( i\int_{t_N^i}^{t_N^f} dt\,H_I^{(N)}(t)\rt)\
    \Or\rt\ket\rt\ket}{
    \lt\bra\lt\bra \Pexp\lt( i\int_{t_1^i}^{t_1^f} dt\,H_I^{(1)}(t)\rt)
    \ \Pexp\lt( i\int_{t_2^i}^{t_2^f} dt\,H_I^{(2)}(t)\rt) \cdots
    \Pexp\lt( i\int_{t_N^i}^{t_N^f} dt\,H_I^{(N)}(t)\rt)
    \rt\ket\rt\ket} \n
&&    =
    \lt\bra\lt\bra \Pexp\lt( i\int_{t_1^i}^{t_1^f} \hspace{-0.3cm}
        dt\,H_I^{(1)}(t)\rt)
    \Pexp\lt( i\int_{t_2^i}^{t_2^f} \hspace{-0.3cm} dt\,H_I^{(2)}(t)\rt) \cdots
    \Pexp\lt( i\int_{t_N^i}^{t_N^f} \hspace{-0.3cm} dt\,H_I^{(N)}(t)\rt)
    \Or\rt\ket\rt\ket_{\rm connected}~. \no
\eeqa
On the right-hand side, the operators $H_I^{(1)}(t_i)$'s,
$H_I^{(2)}(t_i)$'s, ..., $H_I^{(N)}(t_i)$'s are all connected to
$\Or$.

\section{Proof of factorization}

\label{factorization}

Consider a product of operators of the type
\[
    \vec{J}_1(x+x_1)\cdot \vec{S} \,\vec{J}_2(x+x_2)\cdot
    \vec{S}\cdots \vec{J}_n(x+x_n)\cdot\vec{S}
\]
where $\vec{J}_{1,2,\ldots}$ can be $\vec{J}_d,\,\vec{J}_x$ or
$\vec{J}_y$. Recall the notation (\ref{freetrace}). There, the
trace is performed over the Hilbert space of the conformal field
theory where the currents $\vec{J}_{1,2,\ldots}$ act and over the
two-dimensional impurity space associated to $\vec{S}$. Insert an
operator which is composed of products of local operators acting
on the CFT Hilbert space tensored with an arbitrary operator on
the impurity space. We will denote it by
\[
    \sum_a \Or^a S_a
\]
where $a=0,1,2,3$ and $S_0$ is the identity $\bf 1$ on the
impurity space. The operators in $\Or^a$ can be at any fixed time
(with respect to the theory $H_0$) and position, and can also be
integrals of such operators over finite time intervals. We the
consider the SU(2) invariant quantity:
\beqa \label{initialexp}
&& \bra\bra \vec{J}_1(x+x_1)\cdot \vec{S} \;\vec{J}_2(x+x_2)\cdot
\vec{S}\cdots \vec{J}_n(x+x_n)\cdot\vec{S} \;\Or^a S_a \ket\ket_0
\n && \qquad\qquad = \bra\bra
J_1^{i_1}(x+x_1)\,J_2^{i_1}(x+x_2)\cdots J_n^{i_n}(x+x_n) \Or^a
\ket\ket_0\;\bra\bra S_{i_1} S_{i_2} \cdots S_{i_n} S_a
\ket\ket_0~.
\eeqa
The first factor on the right-hand side factorizes as
$|x|\to\infty$ because of the locality of the operators and
because the correlation function is evaluated in a unitary quantum
field theory:

\beq
  \bra\bra J_1^{i_1}(x+x_1)\,J_2^{i_1}(x+x_2)\cdots J_n^{i_n}(x+x_n)
\Or^a \ket\ket_0 \;\; \stackrel{|x|\to\infty}\sim \;\;
    \bra\bra J_1^{i_1}(x_1)\,J_2^{i_2}(x_2)\cdots
        J_n^{i_n}(x_n)\ket\ket_0\, \bra\bra \Or^a \ket\ket_0 \no
\eeq
This field theoretic factorization induces a corresponding
factorization in the spin space, as we proceed to show.  The
expression $\bra\bra J_1^{i_1}(x_1)\,J_2^{i_2}(x_2)\cdots
J_n^{i_n}(x_n)\ket\ket_0$ is a tensor in the product space of $n$
copies of the fundamental representation of rhe rotation group
$O(3)$ and the only way to form a rotational invariant with it is
to multiply it by a corresponding product of matrices $S_{i_1}
S_{i_2} \cdots S_{i_n}$, i.e.  $\bra\bra
J_1^{i_1}(x_1)\,J_2^{i_1}(x_2)\cdots
J_n^{i_n}(x_n)\ket\ket_0\,S_{i_1} S_{i_2} \cdots S_{i_n} \propto
{\bf 1}~. $
Hence, the product $S_{i_1} S_{i_2} \cdots S_{i_n}$ also
factorizes, and we have
\beq
 \bra\bra
J_1^{i_1}(x_1)\,J_2^{i_1}(x_2)\cdots J_n^{i_n}(x_n)\ket\ket_0\,
\bra\bra \Or^a \ket\ket_0\;\bra\bra S_{i_1} S_{i_2} \cdots
S_{i_n}\ket\ket_0\,\bra\bra S_a \ket\ket_0
\eeq
The factorization property then follows:
\beqa &&
\bra\bra \vec{J}_1(x+x_1)\cdot \vec{S} \;\vec{J}_2(x+x_2)\cdot
\vec{S}\cdots \vec{J}_n(x+x_n)\cdot\vec{S}\; \Or^a S_a \ket\ket_0
\n && \qquad\qquad \stackrel{|x|\to\infty}\sim \bra\bra
\vec{J}_1(x+x_1)\cdot \vec{S} \,\vec{J}_2(x+x_2)\cdot
\vec{S}\cdots \vec{J}_n(x+x_n)\cdot\vec{S}\ket\ket_0\;\bra\bra
\Or^a S_a \ket\ket_0~.
\eeqa

The operator $H_I$ is a linear combinations of operators of the
type $\vec{J}\cdot\vec{S}$. The operators $\vec{J}$ are
right-moving fields. Hence when $H_I$ is evolved in time with
$H_0$ for a time $t$, it becomes a linear combinations of
$\vec{J}\cdot\vec{S}$'s at position $-t$. This shows the
factorization of juxtaposed $H_I^{(0)}(t)$'s as $|t|\to\infty$
when they are evaluated inside a trace with operators at fixed
time and position on the right. The same proof applies if such
operator insertion is put on the left of juxtaposed
$H_I^{(0)}(t)$'s. This completes the proof.

\section{Callan-Symanzik equation for matrix elements}

\label{appcalsym}

In this appendix, we will justify equation (\ref{CSoper}).
Consider the finite-temperature average, in the theory described
by $H|_{V=0}$, of the regularized current operator
$\cur_\Lambda=\lambda (\vec{J}_x)_\Lambda(0)\cdot\vec{S}$ with
insertions of creation and anihilation operators of the
Hamiltonian: $A_i^\dag(p)$ and $A_i(p)$. More precisely these are
creation and annihilation operators for eigenstates of $H|_{V=0}$
with energy $p$, corresponding to the massless particles
``naturally'' associated to the local operator $J_i(x)$, where
$J_i$ is any of the ten operators $\vec{J}_d$, $\vec{J}_x$,
$\vec{J}_y$, $J_z$. Thus consider,
\beq\label{equilaver}
    \frc{\Tr\lt(e^{-\beta H|_{V=0}}
    A_{i_1}(p_{1})\cdots A_{i_m}(p_m) \cur_\Lambda
    A^\dag_{i_{m+1}}(p_{m+1}) \cdots A^\dag_{i_{m+n}}(p_{m+n}) \rt)}{\Tr\lt(e^{-\beta
    H|_{V=0}}\rt)}~.
\eeq
Since the creation and annihilation operators are eigenoperators
of the Hamiltonian, the quantity (\ref{equilaver}) satisfies the
Callan-Symanzik equation (\ref{CS}) (that is, in (\ref{CS}) we can
replace the steady-state average of the current by this quantity).

But we showed that the interacting density matrix $e^{-\beta
H|_{V=0}}$ can be obtained from the free one $e^{-\beta H_0}$ by
evolving it for an infinite time (\ref{equil}). In much the same
way, the interacting creation and annihilation operators can be
heuristically written in terms of the mode operators associated to
the current algebra operators:
\[
    A^\dag_i(p) = S(0,-\infty) J^\dag_i(p) S(-\infty,0)~,\quad
    A_i(p) = S(0,-\infty) J_i(p) S(-\infty,0)
\]
(where $S(t_1,t_2)$ is defined in (\ref{S})). Here, $A_i$ is any
of the 10 operators $\vec{A}_d$, $\vec{A}_x$, $\vec{A}_y$ or
$A_z$. Indeed, the operators $A_i^\dag(p)$ and $A_i(p)$ written in
this way heuristically satisfy the canonical commutation relations
amongst them, and the appropriate commutation relations with
Hamiltonian $H|_{V=0}$, written as $S(0,-\infty) H_0 S(-\infty,0)$
(they are eigneoperators of $H|_{V=0}$). Hence, the quantity
(\ref{equilaver}) can be written
\beq\label{eafree}
    \frc{\Tr\lt(e^{-\beta H_0}
    J_{i_1}(p_{1})\cdots J_{i_m}(p_m) S(-\infty,0) \cur_\Lambda
    S(0,-\infty)
    J^\dag_{i_{m+1}}(p_{m+1}) \cdots J^\dag_{i_{m+n}}(p_{m+n}) \rt)}{\Tr\lt(e^{-\beta
    H_0}\rt)}~.
\eeq
Taking the zero-temperature limit $\beta\to\infty$, this quantity
reproduces all matrix elements of the current (see
eqn(\ref{barcur}))
 $\bar\cur_\Lambda =
S(-\infty,0)\cur_\Lambda S(0,-\infty)$  on the Hilbert space of
the free theory ${\cal H}_0$, which proves (\ref{CSoper}).

In fact, to be more precise, when going more from (\ref{eafree})
to (\ref{equilaver}), one needs to modify slightly (\ref{eafree}).
First, the mode operators $J_i^\dag(p)$, $J_i(p)$ should be
replaced by appropriate wave packets $\tilde{J}_i^\dag(p)$,
$\tilde{J}_i(p)$, obtained by integrating the associated currents
$J_i(x)$ (times the oscillating exponential $e^{-ipx}$) over space
with a kernel vanishing (exponentially, say) as $|x|\gg l$ for
some wave-packet extent $l$. Then, the evolution operator
$S(-\infty,0)$ should be replaced by $S(t_0,0)$ and
$S(0,\-\infty)$ by $S(0,t_0')$, and these two evolution operators
should be put into the trace in the denominator as well,
 in the order in which they appear in the numerator. By
using arguments involving the vanishing of connected correlation
functions, it should be observed that the limits $|t_0|\gg l,
\beta$ and $|t_0'|\gg l, \beta$ exist independently. Bringing both
evolution operators around the density matrix $e^{-\beta H_0}$,
one should then use the steps involved in proving (\ref{equil}).
The result, up to vanishing contributions as $t_0,t_0'\gg
l,\beta$, is (\ref{equilaver}), with operators $A^\dag_i(p)$ and
$A_i(p)$ replaced respectively by
\[
    \tilde{A}^\dag_i(p) = S(0,t_0'+i\beta) \tilde{J}^\dag_i(p) S(t_0,0)~,\quad
    \tilde{A}_i(p) = S(0,t_0'+i\beta) \tilde{J}_i(p) S(t_0,0)~.
\]
These are the operators that should correspond to asymptotic
states, in the limit $|t_0|,|t_0'| \gg l\gg \beta$ with
$t_0=t_0'<0$. In this limit, one indeed recovers (\ref{eafree}).

\section{Asymptotic behavior of the function $P(w)$}

\label{appasympP}

In this appendix we evaluate the asymptotic behavior of the function $P(w)$
(\ref{Pw}). The large $w$ behavior is easy to
obtain; the first term inside the parenthesis (on the right-hand
side of the second equation in (\ref{Pw})) disappears, and the
second term $-(p-1)/(e^{w(p-1)}-1)$ becomes $(p-1)\Theta(p<1)$: it
is non-zero only for $p<1$. Taking these contributions into
account as well as the last term $e^{-p^2}$, the result is:
\beq
    P(\infty) = 1-\frc{\gamma}2~.
\eeq
In fact, we can also obtain the next term in the large $w$
expansion. The next contributions can be written in the following
way, by expanding the integrand:
\[
    \int_{0}^{\infty} \frc{d p}p \sum_{n=1}^\infty \lt((p+1)e^{-nw(p+1)}
    +\Theta(p<1)(p-1)e^{nw(p-1)}-\Theta(p>1)(p-1)e^{nw(1-p)}
    \rt)~.
\]
Interchanging integration and summation (this is valid because the
integration variable $p$ is always kept in a region where the
expansion of the integrand is convergent), the integral of every
term of the sum can be expressed in terms of the exponential
integrals. Every term can then be expanded at large $w$, giving
\[
    \sum_{n=1}^\infty \lt( -\frc2{n^2w^2} -\frc{12}{n^4w^4} - \frc{240}{n^6w^6} - \ldots \rt)~.
\]
Now every term can be re-summed, and we get the
large-$w$ asymptotic expansion (\ref{Plargew}).

The small $w$ behavior is more subtle. The integral can be divided
into two parts:
\beq
    P(w) = \lt(\int_0^1 + \int_1^\infty\rt) \frc{d p}p\lt(\frc{p+1}{e^{w(p+1)}-1} -
    \frc{p-1}{e^{w(p-1)}-1} + e^{-p^2}\rt)~.
\eeq
The first integral can be evaluated by expanding the integrand in
small $w$. The integrand goes as $-1 + e^{-p^2} + O(w)$ at small
$w$, so the first integral is convergent at $w\to0$, and in fact
gives an expansion in Taylor series in $w$. In order to obtain the
divergent part in $w$, we need only consider the second integral,
and we can forget about the term $e^{-p^2}$. Make the
transformation of variable $p\to p/w$:
\[
    \int_{w}^\infty \frc{d p}p\lt(\frc{p/w+1}{e^{p+w}-1} -
    \frc{p/w-1}{e^{p-w}-1}\rt)~.
\]
The integrand can then be expanded in $w$: this gives a Taylor
series in $w^2$ starting with power 0. Each term of this Taylor
series gives a convergent integral: the asymptotic behavior of
each term at $p\to\infty$ is exponentially decreasing. Moreover,
each term, except for the very first one, has a behavior like
$p^0$ as $p\to0$, so that at each order in $w^2$, except at the
zeroth order, we can evaluate the integral. We thus obtain a
Taylor series in $w^2$. At the zeroth order, we have
\[
    \int_{w}^\infty \frc{d
    p}p\lt(-2\, \frc{(p-1)\,e^p+1}{(e^p-1)^2}\rt)~.
\]
We can write it as
\[
    \int_{w}^\infty \frc{d
    p}p\lt(-2\,\frc{(p-1)\,e^p+1}{(e^p-1)^2} + \Theta(p<1)\rt) +
    \int_{w}^\infty \frc{d p}p (-\Theta(p<1))~.
\]
The first integral is convergent as $w\to0$: it has again a Taylor
expansion in $w$. The second integral is easy to evaluate, and
gives $\ln(w)$. Hence,
\beq
    P(w) \sim \ln(w) \quad\mbox{as}\quad w\to0~.
\eeq
In fact, we can gather the previous missing parts to get the
constant term:
\beq
    \int_0^1 (-1) \,dp + \int_0^\infty e^{-p^2} + \int_{0}^\infty \frc{d
    p}p\lt(-2\,\frc{(p-1)\,e^p+1}{(e^p-1)^2} + \Theta(p<1)\rt)~.
\eeq
This gives
\beq
    \frc{\sqrt{\pi}}2 +\gamma-\ln(2\pi)
    = -0.374434476...
\eeq
so that we get (\ref{Psmallw}).

\section{Integrals for the two-loop calculations}

\label{appintegtwoloop}

The integrals, as they enter in (\ref{curI2}), are
\beqa
    I_4 &=& \frc{3iV}8 \int
        dp\,R(p)\lt(-i\pi\delta(p) + \prin\frc1{p}\rt)
        \int dq\,R(q)\lt(-i\pi\delta(p+q) + \prin\frc1{p+q}\rt)
        \times\no\\
        && \qquad\quad
        \int dr\,R(r)\lt(-i\pi\delta(p+q+r) +
        \prin\frc1{p+q+r}\rt)\,R(-p-q-r)
        \times\no\\
        && \qquad\qquad\quad
        \Big(-2g(p+q)(-f(p+q+r)+f(r)) + g(q)(-f(p+q+r)+f(p+r)-f(q+r)+f(r)\Big)\n
    I_5 &=& -\frc{3iV}8 \int
        dp\,R(p)\lt(-i\pi\delta(p) + \prin\frc1{p}\rt)
        \int dq\,R(q)\lt(-i\pi\delta(p+q) + \prin\frc1{p+q}\rt)
        \times\no\\
        && \qquad\quad
        \int dr\,R(r)\lt(-i\pi\delta(p+q+r) +
        \prin\frc1{p+q+r}\rt)\,R(-p-q-r)
        \times\no\\
        && \qquad\qquad\quad
        \Big(-2g(r)(f(p+q)-f(p+q+r)) + g(r)(f(q)-f(q+r)) + g(p+r)(f(q)-f(p+q+r)\Big)\n
    && -\frc{3iV}4 \int
        dp\,R(p)^2\lt(\prin\frc1{p}\rt)
        \int dq\,R(q)^2\lt(-i\pi\delta(p+q) + \prin\frc1{p+q}\rt)
        \lt(-i\pi\delta(q) +
        \prin\frc1{q}\rt)\,
        p f(q)\n
    && -\frc{3iV}4 \int
        dp\,R(p)^2\lt(-i\pi\delta(p)\rt)
        \int dq\,R(q)^2\lt(i\pi\delta'(q) + \prin\frc1{q^2}\rt)
        p f(q)\n
    I_6 &=& \frc{3iV}8 \int
        dp\,R(p)\lt(-i\pi\delta(p) + \prin\frc1{p}\rt)
        \int dq\,R(q)\lt(-i\pi\delta(p+q) + \prin\frc1{p+q}\rt)
        \times\no\\
        && \qquad\quad
        \int dr\,R(r)\lt(-i\pi\delta(p+q+r) +
        \prin\frc1{p+q+r}\rt)\,R(-p-q-r)
        \times\no\\
        && \qquad\qquad\quad
        \Big(2g(r)(f(p+q)-f(p+q+r)) + g(q)(-f(p+q+r)+f(p+r)) + 4\Big)\n
    &&  + \frc{3iV}2 \int
        dp\,R(p)^2\lt(i\pi\delta'(p) + \prin\frc1{p^2}\rt)
        \int dq\,R(q)^2\lt(-i\pi\delta(p+q) + \prin\frc1{p+q}\rt)
        q g(q)\n
    I_7 &=& -\frc{3iV}8 \int
        dp\,R(p)\lt(-i\pi\delta(p) + \prin\frc1{p}\rt)
        \int dq\,R(q)\lt(-i\pi\delta(p+q) + \prin\frc1{p+q}\rt)
        \times\no\\
        && \qquad\quad
        \int dr\,R(r)\lt(-i\pi\delta(p+q+r) +
        \prin\frc1{p+q+r}\rt)\,R(-p-q-r)
        \times\no\\
        && \qquad\qquad\quad
        \Big(2g(p+q)(-f(p+q+r)+f(r)) + g(p+r)(f(q)-f(p+q+r)) -g(q+r)(f(q)+f(r)) + 1 \Big)\n
    &&  - \frc{3iV}4 \int
        dp\,R(p)^2\lt(\prin\frc1{p}\rt)
        \int dq\,R(q)^2\lt(-i\pi\delta(p+q) + \prin\frc1{p+q}\rt)
        \lt(-i\pi\delta(q) +
        \prin\frc1{q}\rt)\,
        p f(q) \n
    &&  - \frc{3iV}4 \int
        dp\,R(p)^2\lt(-i\pi\delta(p)\rt)
        \int dq\,R(q)^2\lt(i\pi\delta'(q) + \prin\frc1{q^2}\rt) \,
        p f(q)
\eeqa
where
\beqa
    R(p) &=& R_\Lambda(Vp) = e^{-\frc{V^2p^2}{2\Lambda^2}}~,\n
    g(p) &=& \frc{1+e^{-wp}}{1-e^{-wp}} ~.
\eeqa
Recall that $w=\beta V$. The function $f(p)$ is as in (\ref{fp}),
and is related to $g(p)$ by
\beq
    f(p) = (p+1)g(p+1) - (p-1)g(p-1)~.
\eeq

\section{Finite contributions of the two-loop results}

\label{appfinitetwoloop}

They are given by
\beqa
    [I_4 + I_5 + I_6]_{{\rm finite}} &=&
    6\pi V \int_0^\infty
        \frc{d q}q \int_0^q
        \frc{dr}r\,
        \lt(g_c(q+r)(f(q)+f(r))
         - g_c(q-r)(f(q)-f(r)) \rt) \n
&&
    -\frc{3\pi V}2 \int_0^\infty\frc{d q}q \int_0^q\frc{d r }r\,
        \Big[ g_c(q-r)f_c(q-r)-g_c(q+r)f_c(q+r)\Big] \n
&&    +\frc{3\pi V}2 \int_0^\infty\frc{d q}q \int_0^q
        \frc{dr}r\,
        \lt[ h_c(q+r)-\frc{q}{q-r}h_c(q-r)e^{-r^2}\rt] \n
&&    +\frc{3\pi V}2 \int_0^\infty\frc{d q d r}{q
        r}\,(g_c(r)+e^{-r^2})
        (2f(q)-f(q+r)-f(q-r)) \n
    &&
    -\frc{3\pi V}2 \int_0^\infty\frc{d q d r}{q
        r}\,(1-e^{-r^2})\,(f_c(q+r)+f_c(q-r))
        \n
&&
    -12\pi V \int_0^\infty \frc{dr}r\lt(\frc{\gamma}2 + \ln r\rt)
        f_c(r)\n
&&    -\frc{3\pi V}2 \int_0^\infty \frc{d r}r
        \lt(\frc{\gamma}2+\ln r\rt)\,h_c(r) \n
    &&
    -\frc{3\pi V}2 \int_0^\infty \frc{d q}q f_c(q) \n
    && +\frc{3\pi^3 V}2 \lt(\frc{3}w G(w)-1\rt) \n
    &&
     +3\pi V \lt(u(1-i)+u(1+i)
    -u\lt(\frc{1-i}2\rt) -u\lt(\frc{1+i}2\rt)\rt)
    \n
    &&
    -\frc{\pi^3V}{24}
\eeqa
and
\beqa
    [I_7]_{{\rm finite}} &=&
    \frc{3\pi V}2 \int_0^\infty
        \frc{d q}q \int_0^q \frc{d r}{ r}\,
        \Big[
        g_c(q-r)(f(q)-f(r))
        - g_c(q+r)(f(q)+f(r)) \Big] \n &&
    +\frc{3\pi V}2 \int_0^\infty\frc{d q}q \int_0^q\frc{d r }r\,
        \Big[ g_c(q-r)f_c(q-r)-g_c(q+r)f_c(q+r)\Big] \n
&&    -\frc{3\pi V}2 \int_0^\infty\frc{d q}q \int_0^q
        \frc{dr}r\,
        \lt[ h_c(q+r)-\frc{q}{q-r}h_c(q-r)e^{-r^2}\rt] \n
    &&  - 6\pi V
        \int_0^\infty \frc{d q}q \int_0^q \frc{d r}{
        r}\,(1-e^{-r^2})(1-e^{-q^2})
        f_c(q+r) \n
&&
    +3\pi V \int_0^\infty \frc{dr}r\lt(\frc{\gamma}2 + \ln r\rt)
        f_c(r)\n &&
    +\frc{3\pi V}2 \int_0^\infty \frc{d r}r
        \lt(\frc{\gamma}2+\ln r\rt)\,h_c(r) \n
    &&
    -\frc{3\pi V}2 \int_0^\infty \frc{d q}q f_c(q) \n
    &&
    +6\pi
    V\lt[u(i-1)+u(-i-1)+u\lt(\frc{i-1}2\rt)+u\lt(\frc{-i-1}2\rt)\rt]\n
    &&
    + \frc{3\pi V\ln2}2
        + \frc{5\pi^3 V}{4}~.
\eeqa
Here, we have
\beqa
    g(p) &=& \frc{1+e^{-wp}}{1-e^{-wp}} \n
    f(p) &=& (p+1)g(p+1) - (p-1)g(p-1) \n
    g_c(p) &=& g(p) - {\rm sign}(p) \n
    f_c(p) &=& f(p) - 2{\rm sign}(p)(1-e^{-p^2}) \n
    h_c(p) &=& f_c(p) + (1-e^{-p^2})g_c(p) \n
    G(w) &=& \frc{\sinh(w)-\frc{w}2}{\sinh^2\lt(\frc{w}2\rt)} \n
    u(x) &=& -{\rm dilog}(1-x)
\eeqa
where $w=V/T$.

\end{document}